# High-Resolution Observations of Bright Boulders on Asteroid Ryugu: 1. Size Frequency Distribution and Morphology

Chiho Sugimoto[1], Eri Tatsumi[1,2], Yuichiro Cho[1], Tomokatsu Morota[1], Rie Honda[3], Shingo Kameda[4], Yosuhiro Yokota[3,5], Koki Yumoto[1], Minami Aoki[1], Daniella N. DellaGiustina[6,7], Tatsuhiro Michikami[8], Takahiro Hiroi[9], Deborah L. Domingue[10], Patrick Michel[11], Stefan E. Schröder[12], Tomoki Nakamura[13], Manabu Yamada[14], Naoya Sakatani[4], Toru Kouyama[15], Chikatoshi Honda[16], Masahiko Hayakawa[5], Moe Matsuoka[5], Hidehiko Suzuki[17], Kazuo Yoshioka[1], Kazunori Ogawa[18], Hirotaka Sawada[5], Masahiko Arakawa[15], Takanao Saiki[5], Hiroshi Imamura[5], Yasuhiko Takagi[19], Hajime Yano[5], Kei Shirai[15], Chisato Okamoto[15†], Yuichi Tsuda[5], Satoru Nakazawa[5], Yuichi Iijima[5†], and Seiji Sugita[1,14,*]

[1] The University of Tokyo, Tokyo 113-0033, Japan.

[2] Instituto de Astrofísica de Canarias (IAC), 38205 La Laguna, Tenerife, Spain.

[3] Kochi University, Kochi 780-8520, Japan.

[4] Rikkyo University, Tokyo 171-8501, Japan.

[5] Institute of Space and Astronautical Science (ISAS), Japan Aerospace Exploration Agency (JAXA), Sagamihara 252-5210, Japan.

[6] Lunar and Planetary Laboratory, University of Arizona, Tucson, AZ 85721, USA.

[7] Department of Geosciences, University of Arizona, Tucson, AZ 85721, USA.

[8] Faculty of Engineering, Kindai University, Higashi-Hiroshima 739-2116, Japan.

[9] Department of Earth, Environmental and Planetary Sciences, Brown University, Providence, RI 02912, USA.

[10] Planetary Science Institute, Tucson, AZ 85719, USA.

[11] Université Côte d'Azur, Observatoire de la Côte d'Azur, Centre National de le Recherche Scientifique, Laboratoire Lagrange, 06304 Nice, France.

[12] German Aerospace Center (DLR), Institute of Planetary Research, 12489 Berlin, Germany.

[13] Department of Earth Science, Tohoku University, Sendai 980-8578, Japan.

[14] Planetary Exploration Research Center, Chiba Institute of Technology, Narashino 275-0016, Japan.

[15] National Institute of Advanced Industrial Science and Technology, Tokyo 135-0064 Japan.

[16] University of Aizu, Aizu-Wakamatsu 965-8580, Japan.

[17] Meiji University, Kawasaki 214-8571, Japan.

[18] Kobe University, Kobe 657-8501, Japan.

[19] Aichi Toho University, Nagoya 465-8515, Japan

[†] Deceased.

[*] Corresponding author.

E-mail address: sugita@eps.s.u-tokyo.ac.jp

*Abstract:* The near-Earth asteroid (162173) Ryugu displays a Cb-type average spectrum and a very low average normal albedo of 0.04. Although the majority of boulders on Ryugu have reflectance spectra and albedo similar to the Ryugu average, a small fraction of boulders exhibit anomalously high albedo and distinctively different spectra. A previous study (Tatsumi et al., 2021 *Nature Astronomy*, 5, doi:10.1038/s41550-020-1179-z) based on the 2.7-km observations and a series of low-altitude (down to 68 m) descent observations conducted prior to the first touchdown have shown that the spectra of these anomalous boulders can be classified into two distinct groups corresponding to S and C type asteroids. The former originate most likely from an impactor that collided with Ryugu's parent body, whereas the latter may be from portions of Ryugu's parent body that experienced a different temperature history than experienced by the majority of boulder materials. In this study, we analyzed images captured after the first touchdown to determine the quantitative properties of these bright boulders on Ryugu. We measured the sizes of more than a thousand bright boulders and characterized the morphologic properties of the largest ones. Analyses revealed many properties of bright boulders important for the evolution of Ryugu and its parent body. First, the size–frequency distributions of S-type and C-type bright boulders follow a power law with exponents of 1.6±1.3 and 3.0±0.7, respectively. Based on these size–frequency distributions, we obtained the ratios of the total volume and surface area of S-type bright boulders to those of average dark boulders on the Ryugu's surface, that is, $7.1^{+6.3}_{-5.0} \times 10^{-6}$ and $1.5^{+3.2}_{-1.2} \times 10^{-6}$, respectively, over the diameter range of 0.3 to 3 m. Similarly, the ratio of the total volume and surface area of C-type bright boulders to those of average dark boulders are $4.4^{+14.0}_{-2.2} \times 10^{-5}$ and $1.3^{+9.8}_{-1.1} \times 10^{-3}$, respectively, at a diameter range of 2 cm to 2 m. Second, the number density of bright boulders inside the artificial crater newly made by the Small Carry-on Impactor (SCI) experiment agrees with the outside number density within a factor of two. Third, many of the bright boulders are embedded in a larger substrate boulder, suggesting that they have experienced mixing and conglomeration with darker fragments on Ryugu's parent body, rather than gently landing on Ryugu during or after its formation by reaccumulation. This observation is consistent with the hypothesis that S-type bright boulders were likely mixed during and/or before a catastrophic disruption. C-type bright boulders embedded in substrate boulders suggests a brecciation process after thermal metamorphism. Furthermore, the embedding of S-type clasts in substrate boulders suggests that brecciation did indeed occur even after a large-scale impact on the parent body. If the brecciation on the Ryugu's parent body occurred over such a long period or over many stages of its evolution, breccias may end up being the dominant constituent materials on Ryugu's parent body. Moreover, the preponderance of breccias may contribute to the globally low thermal inertia of Ryugu.

# 1. Introduction

Recent spacecraft observations have shown that many small asteroids, such as Itokawa, Ryugu, and Bennu, are rubble-pile asteroids (Fujiwara et al., 2006; Watanabe et al., 2019; Lauretta et al., 2019). Owing to their short collisional lifetime, they were likely formed recently (~$10^8$ years) from much larger parent bodies due to impact-induced catastrophic disruption (O'Brien and Greenberg, 2005; Bottke et al., 2005, Michel & Ballouz et al., 2020). These theoretical predictions are largely consistent with the properties of asteroid families revealed via telescopic observations, such as parent body sizes, their spectral types, and collisional timings (Masiero et al., 2015). However, many other important aspects of the collisional evolution of asteroids, from the catastrophic disruption of the parent bodies to the reaccumulation and/or formation of rubble-pile asteroids, are not fully understood thus far. More specifically, projectile types, spectral variety within the pre-impact parent bodies, and brecciation processes on parent bodies, particularly for low-albedo carbonaceous chondrites (CCs), most of which are known to be breccias (e.g., Bischoff et al., 2006), are not fully understood. Understanding such specific properties of individual collision events and subsequent geologic processes requires highly detailed observations of rubble-pile asteroids.

The Hayabusa2 mission observed a C-type rubble-pile asteroid in detail (Watanabe et al., 2017). In particular, the initial observation of Ryugu using the Optical Navigation Camera telescope (ONC-T) onboard Hayabusa2 revealed that the color of Ryugu's surface is very uniform and extremely dark, and that the average spectrum is similar to aqueously altered and subsequently thermally metamorphosed carbonaceous chondrites (Sugita et al., 2019). The geometric albedo of Ryugu is 4.0% at 0.55 $\mu$m (Tatsumi et al., 2020), one of the darkest bodies in the Solar System. Thus, anomalous materials that deviate from the average reflectance on Ryugu are expected to stand out as bright spots. Hayabusa2's initial observations at an altitude of ~20 km (image resolution of ~2.1 m/pixel) revealed the presence of bright spots (Sugita et al., 2019). Subsequently, Tatsumi et al. (2021) discovered many bright spots based on higher-resolution (~0.29 m/pixel) images captured during the hovering operation after the deployment of the CNES-DLR MASCOT lander at ~2.7 km in altitude and images (7.3 mm/pixel) acquired during the MINERVA-II hopper deployment operation down to ~68 m in altitude. With such datasets, it was found that the bright spots are boulders. It is noteworthy that many small depressions (i.e., mini-craters) are observed on Bennu and that their colors are either similar to the surrounding surfaces or darker, not brighter (Ballouz et a. 2020).

Analyses of 21 bright boulders on Ryugu by Tatsumi et al. (2021) revealed many of their important, distinctive properties. The main findings can be summarized as follows: (1) There are two spectral types: C/X and S. (2) The spectra of S-type bright boulders at visible–near-infrared wavelengths are consistent with ordinary chondrites (OCs). (3) The size and number density of S-type bright boulders indicate that these boulders are too large and too numerous to have been accreted on Ryugu after its formation to its current size and mass. Furthermore, because the mass of impactor(s) that catastrophically disrupted (CD) Ryugu's parent body would be much greater than impactors that did not, the likelihood that exogenic fragments are from CD impactor(s) is greater than non-CD impactors. Consequently, S-type bright boulders may be remnants of the impactor that catastrophically disrupted Ryugu's parent body and that contributed to the reaccumlation process that formed Ryugu (Michel and Ballouz et al., 2020). (4) The visible spectra of C-/X-type bright boulders are consistent with heated CCs. Because of the similarity to CCs, we call these boulders C-type hereafter. (5) The size frequency distribution (SFD) of bright boulders is similar to that of other large (1–100 m) boulders with an average brightness similar to Ryugu's surface (Michikami et al., 2019); both have a power-law distribution index of ~2. Here, it is noted that some "bright boulders" are smaller than the geology definition (i.e., rocks greater than 256 mm) of boulder. For simplicity, however, we call pebble- and cobble-size objects also "bright boulders" in this study and our companion (Sugimoto et al., 2021), which is hereinafter referred to as Paper 2, in accordance with the use of the term by Tatsumi et al. (2021).

Tatsumi et al. (2021) further characterized the properties of these bright boulders in more detail. Observations by the near-infrared spectrometer NIRS3 suggest that the S-type bright boulders have a very weak or no 2-$\mu$m absorption. This is more consistent with OCs than with the Howardite–Eucrite–Diogenite meteorites, even after considering possible contamination from background surface spectra. This finding is particularly important for understanding Ryugu's collisional evolution because bright boulders found on a similarly dark rubble-pile asteroid, Bennu, were identified to contain anhydrous silicates with a 2-$\mu$m absorption features not observed in the Ryugu spectra (DellaGiustina et al., 2021). The different spectral properties of exogenic bright boulders on these rubble pile asteroids suggest that their parent bodies might have been struck by projectiles of different compositions, supporting that they experienced different collisional evolutions.

Tatsumi et al. (2021) found that most of C-type bright boulders show spectra similar to the surface of Ryugu and that a few bright boulders in this group show spectra with much stronger UV absorption and larger spectral slope. Comparison between these C-type bright boulders and heated CCs indicates that the observed color variation of C-type bright boulders is similar in range and pattern to those of heated Murchison (CM2) and heated Ivuna (CI) meteorites.

However, other important properties of bright boulders, such as their morphologies, occurrence and distribution, or details of their color variation, have not yet been investigated. These properties are important for understanding their origins. Furthermore, no bright boulder at higher latitudes has been found, but we underline that high-resolution search for bright boulders have been conducted only at low latitudes thus far. Thus, the spatial distribution of bright boulders on Ryugu has not been well-constrained. Hayabusa2 conducted many other low-altitude observations covering wider areas on Ryugu than the data analyzed by Tatsumi et al. (2021). Consequently, in this study, we conducted a series of analyses of bright boulders on Ryugu using high-resolution (< 10 cm/pixel) images captured by ONC-T at lower altitudes (< 1 km) and with wider surface area coverage at higher latitudes.

High-resolution observations allow us to investigate detailed morphological properties of bright boulders to constrain their origin. Through such imagery dataset, brecciated boulders composed of fragments with different brightness were discovered on Ryugu (Sugita et al., 2019). The incorporation of bright boulders within boulders corresponding to breccia suggests that materials of different composition, including exogenic ones, have congregated and subsequently experienced cementation processes, which are difficult to produce on small rubble-pile bodies. Such incorporation rather suggests mixing through a violent enough process during/before the catastrophic disruption of Ryugu's parent body. Furthermore, to examine whether bright boulders exist only on the surface of Ryugu or whether they also exist in the interior of Ryugu, we investigate images around the artificial crater created by a small carry-on impactor (SCI) onboard Hayabusa2 (Arakawa et al., 2020).

Although high-resolution images also allow us to investigate the spectral variations within bright boulders, this is beyond the scope of this study and is explored in a separate contribution (Sugimoto et al., 2021), which is hereinafter referred to as Paper 2. The remainder of this paper is organized as follows. We discuss the data and analysis method in Section 2, spatial and size frequency distributions in section 3, morphological analysis results in Section 4, geological implications for the evolution of Ryugu and its parent body in Section 5, and conclusions in Section 6.

## 2. Method

To investigate the size frequency and detailed morphology of bright boulders, we used v-band images. In this section, we first discuss the v-band image data reduction in Section 2.1 and then the SFD measurements in Section 2.2.

**2.1 Datasets and processing**

We used v-band (0.55 $\mu$m) images captured by the telescopic Optical Navigation Camera (ONC-T) onboard Hayabusa2 (Kameda et al., 2017; Sugita et al., 2019) for the size–frequency analysis and detailed morphological analysis of bright boulders. We analyzed the data used in Tatsumi et al. (2021) as well as those obtained later in the mission. The former was obtained during hovering operations after MASCOT deployment at ~2.7 km in altitude (~0.29 m/pixel) on 3 to 4 October 2018. The latter includes two sets of high-resolution (~0.18 m/pixel) images captured at ~1.7 km in altitude during the artificial crater search scanning operations before (CRA1) and after (CRA2) the SCI operations (Arakawa et al., 2020) conducted on 21 March and 25 April 2019, respectively. The latter also includes descent observation images (≥ 3.8 mm/pixel) down to ~35 m altitude. Those descent observations include images acquired during the first touchdown rehearsal (TD1-R1A), the first touchdown (TD1), as well as descent observations (DO-S01) around the second touchdown site and low descent observations (PPTD-TM1B) around the SCI crater conducted on 15 October 2018, 21 February 2019, 8 March 2019, and 13 June 2019, respectively. It is noted that exposure times used for the PPTD-TM1B descending sequence are about a half with respect to the other low altitude observations in order to avoid signal saturations for bright boulders seen in other low-altitude observations. For such reason, these images are particularly important for analyzing bright boulders.

For the size–frequency analysis, we analyzed images captured during TD1-R1A, 2.7-km hovering observations, and 1.7-km scanning observations (CRA1). Although 12 images were taken during the CRA1 observations, peripheral areas of some these images are highly distorted. Thus, accurate size measurements of boulders in these images are difficult. Consequently, we decided to use only four central images from the CRA1 observations in the SFD analysis. Illumination and observation geometry was calculated based on the orbital and attitude information. The phase angles $\theta$ of observations during MASCOT hovering is 12.7°±1.8°. Most of this pixel-level variation in phase angle $\theta$ results from the width (±3.1°) of field of view (FOV) of ONC-T. The standard deviation of the image-averaged phase angle is very small (0.13°) because the viewing geometry during MASCOT hovering was nearly constant. In contrast, the spacecraft attitude during CRA1 and CRA2 observations was systematically changed for scanning a large surface area on Ryugu (Arakawa et al. 2020). The phase angle a was 20.3°±6.0° for CRA1 and 26.7°±7.1° for CRA2. The intra-image variation in $\theta$ is the same as the MASCOT hovering series, but the inter-image variation in average $\theta$ is much larger (2.9° for CRA1 and 3.0° for CRA2). Because of large uncertainty in spacecraft trajectories during the low-attitude observations (i.e., TD1-R1A, TD1, DO-S01, and PPTD-TM1B), we could not derive specific illumination conditions for these datasets.

We conducted data processing, such as the removal of stray light, bias, and read-out smear, flat fielding, and I/F conversion based on the calibration data and method described in Suzuki et al. (2018) and Tatsumi et al. (2019). Note that the updated flat fields based on the close encounter images of the asteroids (Kameda et al. 2021; Koyama et al., 2021) were employed.

**2.2 SFD of bright boulders**

As ONC-T has a fixed focal length, its spatial resolution is determined by the altitude of the spacecraft from the surface of Ryugu. Thus, the measurements of bright boulders with sizes over a wide range require observations at altitudes covering a wide range. We analyzed images captured during two descent operations, 2.7-km hovering observations, and 1.7-km scanning observations: TD1 rehearsal, low descent observation around the SCI crater, hovering operation after MASCOT deployment, and the artificial crater search scanning operations before/after the SCI operation. The SFDs obtained for the bright boulders from theses sets of observations were compared with the SFD of general bright boulders studied by Michikami et al. (2019), who measured the SFD of all the boulders including bright boulders in the images taken during the TD1 rehearsal. As for touchdown rehearsal, we analyzed seven images captured at distances from the surface ranging from 1194 to 80 m. We defined objects brighter than 1.5 times the median value of each image as bright boulders. Examples of identified bright boulders in this analysis are shown in Figure B1. This brightness criterion (1.5 times) for bright boulders is the same as that used by Tatsumi et al. (2021) and is higher than the average by $5\sigma$, where $\sigma$ is the standard deviation in surface albedo. Assuming that bright boulders have a circular shape, the diameter $D_{BB}$ (in m) of a bright boulder was calculated from the pixel area $S_{\text{pix}}$ (i.e., the number of pixels that covers bright boulders) using

$$D_{BB} = 2l\sqrt{\frac{S_{pix}}{\pi}}, \qquad (2.2.1)$$

where $l$ (m/pixel) is the pixel scale at the altitude the image was acquired. As errors due to the point-spread function (PSF 1.7 pix) are expected to exist, the relative error in the diameter $D_{BB}$ would be ≲ 66% for very small bright boulders (~3 pix in diameter) and ~20% for larger bright boulders (~10 pix in diameter). Here, the relative error in the size of a bright boulder is given by 2/(*n*+2), when the linear dimension of a BB is *n* pixels. This is because its image size may differ from *n* to *n*+2 pixels depending on where the edges of the BB are located with respect to the the charge-coupled device (CCD) pixel grid. Thus, relative error is 66% and 20% for n=1 and n=10, respectively. This level of precision is still sufficient for our analysis since the SFD is examined in orders of magnitudes and discussed here. The cumulative number distribution of boulder sizes can be expressed by $N(>D) \propto D^{-\alpha}$, where $D$ is the diameter of the bright boulder and $\alpha$ is the power-law index. The boulder distribution in each image is fitted by a power law based on the maximum likelihood method with a goodness-of-fit test based on the Kolmogorov–Smirnov statistics (Clauset et al., 2009), which was implemented in the python library, *powerlaw*, by Alstott et al. (2014).

To investigate the SFD of S-type bright boulders, we used the results of the spectral classification by Tatsumi et al. (2021) and those in Paper 2. Some of the bright boulders observed in the v-band images captured during the 2.7-km hovering observation and 1.7-km scanning observations have been spectroscopically analyzed. The location map of these bright boulders is shown in Fig. 1, and their longitudes and latitudes are listed in Table 1. Note that the sizes listed in this table are not those derived from the Eq. 2.2.1, but those obtained using the size estimation method described in Section 2.2 of Paper 2. The SFD of S-type bright boulders was obtained from the sizes calculated by Eq. 2.2.1. For the SFD of C-type bright boulders, the sizes of bright boulders other than those classified as S type were considered. Three the bright boulders (i.e., C65, C29, and C26) analyzed in this study overlap with the bright boulders (i.e., M7, M19, and M21) discovered by Tatsumi et al. (2021) and the correspondence are given in Table 1.

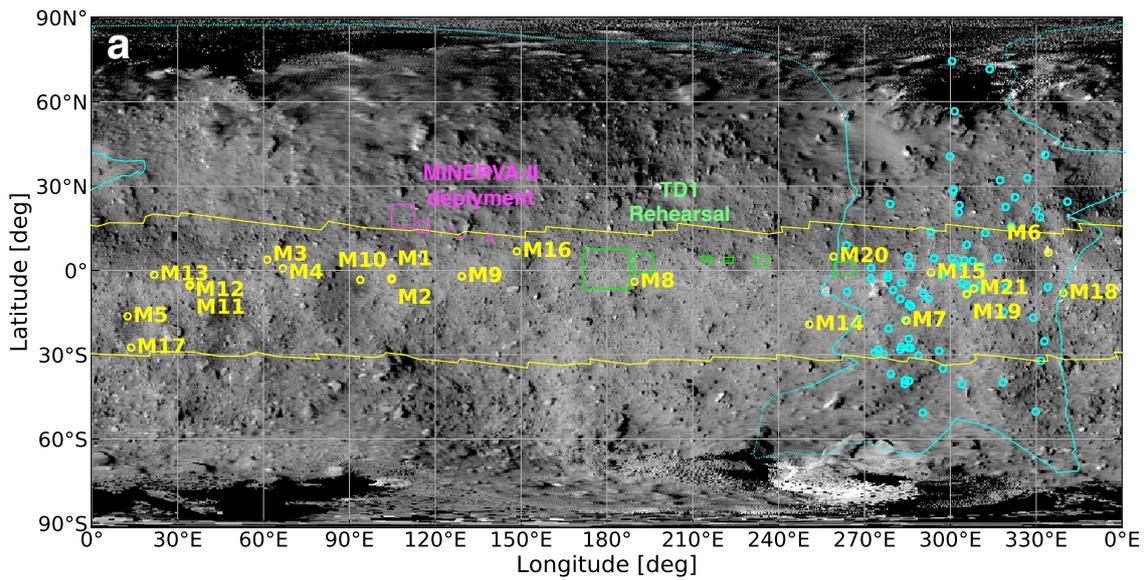

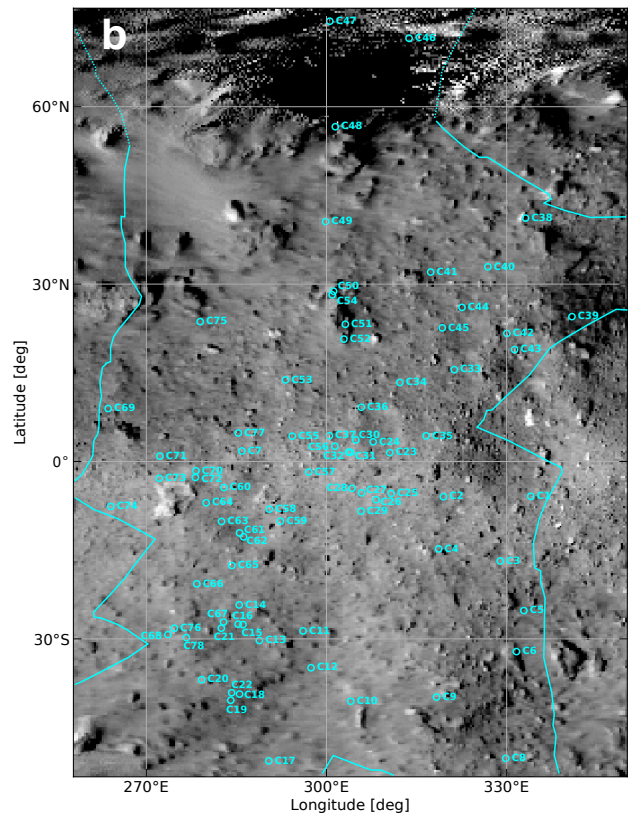

**Fig. 1** Spatial distribution of spectroscopically analyzed bright boulders and coverage of the 2.7-km hovering observations, the 1.7-km scanning observations, and the decent observations. (a) global map and (b) a close-up map around 1.7-km scanning observations. Yellow open circles indicate the locations of M1–M21 bright boulders in Tatsumi et al. (2021) and cyan open circles indicate C1–C79 bright boulders investigated in this study and Paper 2. The yellow boundary indicates the area observed during the 2.7-km hovering

observations (~0.29 m/pixel), and the cyan boundary indicates that of the 1.7-km scanning observations (~0.18 m/pixel). Note that some M-series and C-series boulders overlap. Table 1 presents the details of the relationships. The areas surrounded by magenta and lime thick solid lines indicate the area covered by high-resolution images during the two decent observations: MINERVA-II release operations and TD1 rehearsal.

**Table 1.** Bright boulders observed in CRA1, CRA2, DO-S01, and PPTD-TM1B operations.

| ID | | Latitude [deg]‡ | Longitude [deg]‡ | Diameter [m] $D_{ave}^{\#}$ | $\sigma_{D_l}^{\#}$ | $\sigma_{Du}^{\#}$ | Normal Albedo [%] $A_{normal}^{\dagger}$ | $\sigma_{A_l}^{\dagger}$ | $\sigma_{Au}^{\dagger}$ | Spectral Type§ | Size measurement Method* |
|---|---|---|---|---|---|---|---|---|---|---|---|
| C1 | - | -5.1 | 334.4 | 0.26 | 0.21 | 0.21 | 20.5 | 20.5 | 18.0 | - | B |
| C2 | - | -5.1 | 319.8 | 0.43 | 0.29 | 0.22 | 9.3 | 5.0 | 6.4 | C | B |
| C3 | - | -16.0 | 329.3 | 0.59 | 0.19 | 0.19 | 7.1 | 1.9 | 2.7 | C | B |
| C4 | - | -14.0 | 319.0 | 0.22 | 0.17 | 0.17 | 9.7 | 8.0 | 7.1 | S | B |
| C5 | - | -24.4 | 333.2 | 1.50 | 0.50 | 0.50 | 7.4 | 1.9 | 4.6 | C | A |
| C6 | - | -31.3 | 332.0 | 0.35 | 0.31 | 0.31 | 11.4 | 11.4 | 10.1 | - | B |
| C7 | - | 2.6 | 286.2 | 0.13 | 0.08 | 0.08 | 40.7 | 38.6 | 37.5 | C | B |
| C8 | - | -49.4 | 330.2 | 0.17 | 0.11 | 0.11 | 43.3 | 41.3 | 40.3 | C | B |
| C9 | - | -39.0 | 318.7 | 1.50 | 0.50 | 0.50 | 7.0 | 1.8 | 4.4 | C | A |
| C10 | - | -39.7 | 304.4 | 1.50 | 0.50 | 0.50 | 6.4 | 1.6 | 4.0 | C | A |
| C11 | - | -27.8 | 296.4 | 2.50 | 0.50 | 0.50 | 6.3 | 1.6 | 3.9 | C | A |
| C12 | - | -34.1 | 297.8 | 1.50 | 0.50 | 0.50 | 6.4 | 1.7 | 4.0 | C | A |
| C13 | - | -29.4 | 289.2 | 0.24 | 0.20 | 0.20 | 15.0 | 15.0 | 12.9 | C | B |
| C14 | - | -23.4 | 285.8 | 0.60 | 0.15 | 0.15 | 5.6 | 1.6 | 2.2 | C | B |
| C15 | - | -26.7 | 286.4 | 1.50 | 0.50 | 0.50 | 5.9 | 1.5 | 3.7 | C | A |
| C16 | - | -26.7 | 285.6 | 2.50 | 0.50 | 0.50 | 6.1 | 1.6 | 3.8 | C | A |
| C17 | - | -49.8 | 290.7 | 1.50 | 0.50 | 0.50 | 6.6 | 1.7 | 4.2 | C | A |
| C18 | - | -38.5 | 285.9 | 1.50 | 0.50 | 0.50 | 6.2 | 1.6 | 3.9 | C | A |
| C19 | - | -39.6 | 284.4 | 1.50 | 0.50 | 0.50 | 6.9 | 1.8 | 4.4 | C | A |
| C20 | - | -36.1 | 279.5 | 0.75 | 0.25 | 0.25 | 5.9 | 1.6 | 3.8 | C | A |
| C21 | - | -27.4 | 282.8 | 0.46 | 0.35 | 0.32 | 6.5 | 4.5 | 4.5 | C | B |
| C22 | - | -38.3 | 284.5 | 0.75 | 0.25 | 0.25 | 7.5 | 2.0 | 4.7 | C | A |
| C23 | - | 2.3 | 310.9 | 0.37 | 0.23 | 0.11 | 12.5 | 6.9 | 9.0 | C | B |
| C24 | - | 4.2 | 308.1 | 0.75 | 0.25 | 0.25 | 6.8 | 1.8 | 4.3 | C | A |
| C25 | - | -4.6 | 311.1 | 0.18 | 0.13 | 0.12 | 14.9 | 12.9 | 12.1 | S | B |

#: $D_{ave}^{\#}$, $\sigma_{D_l}^{\#}$, and $\sigma_{Du}^{\#}$ are average diameter of bright boulders, lower error, and upper error, respectively. These were obtained using the size estimation method described in Section 2.2 of Paper 2.
†: $A_{normal}$, $\sigma_{A_l}$, and $\sigma_{Au}$ are average normal albedo, lower error, and upper error, respectively.
‡: The 2σ error of latitude and longitude is 0.4° except for C8, which has greater uncertainty (~1°) because of its location near the edge of the scanned observation region.
§: The spectral type classification is given in Section 3 in Paper 2. Hyphen in this column indicates that spectral type is unclassified.

*: For bright boulders with areas > 9 pixel², we estimated their spectra by averaging the reflectance of the pixels within the area occupied by the boulder for each band (Method A). For bright boulders with areas ≦ 9 pixel², we extracted the spectra of these bright boulders using the following method (Method B). Details of method A and B are given in Section 2.2. in Paper 2.

**Table 1.** (continued)

| ID | | Latitude [deg] | Longitude [deg] | Diameter [m] | | | Normal Albedo [%] | | | Spectral Type§ [deg] | Size measurement Method* [deg] |
|---|---|---|---|---|---|---|---|---|---|---|---|
| | | | | $D_{ave}^{\#}$ | $\sigma_{D_l}^{\#}$ | $\sigma_{Du}^{\#}$ | $A_{normal}^{\dagger}$ | $\sigma_{A_l}^{\dagger}$ | $\sigma_{Au}^{\dagger}$ | | |
| C26 | M21 | -5.6 | 308.6 | 0.75 | 0.25 | 0.25 | 7.2 | 1.9 | 4.6 | C | A |
| C27 | - | -4.5 | 306.2 | 0.21 | 0.17 | 0.17 | 16.8 | 16.8 | 14.5 | C | B |
| C28 | - | -3.7 | 304.6 | 0.43 | 0.29 | 0.24 | 7.6 | 3.8 | 4.6 | C | B |
| C29 | M19 | -7.6 | 306.1 | 0.75 | 0.25 | 0.25 | 7.7 | 2.1 | 4.9 | C | A |
| C30 | - | 4.5 | 305.2 | 0.75 | 0.25 | 0.25 | 8.3 | 2.3 | 5.4 | C | A |
| C31 | - | 2.4 | 304.3 | 0.12 | 0.07 | 0.07 | 33.4 | 26.0 | 29.2 | C | B |
| C32 | - | 2.4 | 304.0 | 0.11 | 0.07 | 0.06 | 23.8 | 18.8 | 20.3 | C | B |
| C33 | - | 16.4 | 321.6 | 0.75 | 0.25 | 0.25 | 7.7 | 2.1 | 4.9 | C | A |
| C34 | - | 14.2 | 312.5 | 0.75 | 0.25 | 0.25 | 10.2 | 2.8 | 6.6 | C | A |
| C35 | - | 5.2 | 316.9 | 0.19 | 0.14 | 0.14 | 17.5 | 16.0 | 14.4 | C | B |
| C36 | - | 10.0 | 306.1 | 3.00 | 0.50 | 0.50 | 6.8 | 1.7 | 4.2 | C | A |
| C37 | - | 5.2 | 300.8 | 2.50 | 0.50 | 0.50 | 6.8 | 1.7 | 4.2 | C | A |
| C38 | - | 42.0 | 333.5 | 6.5 | 0.50 | 0.50 | 7.3 | 1.9 | 4.6 | C | A |
| C39 | - | 25.3 | 341.2 | 0.75 | 0.25 | 0.25 | 7.8 | 2.2 | 5.1 | C | A |
| C40 | - | 33.8 | 327.2 | 0.26 | 0.21 | 0.21 | 14.2 | 13.8 | 11.6 | C | B |
| C41 | - | 32.9 | 317.7 | 0.24 | 0.19 | 0.18 | 19.1 | 17.2 | 14.9 | C | B |
| C42 | - | 22.5 | 330.4 | 0.21 | 0.16 | 0.16 | 12.6 | 10.6 | 9.5 | C | B |
| C43 | - | 19.8 | 331.7 | 0.35 | 0.30 | 0.29 | 10.8 | 8.6 | 7.4 | C | B |
| C44 | - | 26.9 | 322.9 | 0.75 | 0.25 | 0.25 | 7.1 | 1.9 | 4.5 | C | A |
| C45 | - | 23.4 | 319.6 | 1.50 | 0.50 | 0.50 | 7.4 | 2.0 | 4.7 | C | A |
| C46 | - | 72.4 | 314.1 | 2.50 | 0.50 | 0.50 | 6.5 | 1.7 | 4.1 | C | A |
| C47 | - | 75.3 | 300.9 | 6.0 | 0.50 | 0.50 | 6.3 | 1.6 | 4.0 | C | A |
| C48 | - | 57.5 | 301.8 | 4.0 | 0.50 | 0.50 | 6.1 | 1.6 | 3.9 | C | A |
| C49 | - | 41.4 | 300.2 | 0.13 | 0.08 | 0.08 | 31.9 | 26.1 | 27.6 | C | B |
| C50 | - | 29.6 | 301.5 | 7.50 | 0.50 | 0.50 | 6.7 | 1.7 | 4.2 | C | A |

**Table 1.** (continued)

| ID | | Latitude [deg] | Longitude [deg] | Diameter [m] $D_{ave}^{\#}$ | $\sigma_{D_l}^{\#}$ | $\sigma_{Du}^{\#}$ | Normal Albedo [%] $A_{normal}^{\dagger}$ | $\sigma_{A_l}^{\dagger}$ | $\sigma_{Au}^{\dagger}$ | Spectral Type§ [deg] | Size measurement Method* [deg] |
|---|---|---|---|---|---|---|---|---|---|---|---|
| C51 | - | 24.0 | 303.5 | 0.22 | 0.18 | 0.22 | 17.5 | 13.9 | 12.6 | - | B |
| C52 | - | 21.5 | 303.2 | 0.29 | 0.24 | 0.24 | 8.9 | 7.0 | 6.1 | C | B |
| C53 | - | 14.6 | 293.5 | 3.50 | 0.50 | 0.50 | 6.3 | 1.6 | 3.9 | C | A |
| C54 | - | 29.0 | 301.3 | 2.50 | 0.50 | 0.50 | 7.7 | 2.0 | 4.8 | C | A |
| C55 | - | 5.1 | 294.6 | 0.25 | 0.20 | 0.20 | 14.9 | 14.9 | 12.7 | C | B |
| C56 | - | 3.4 | 301.7 | 0.34 | 0.24 | 0.22 | 12.4 | 9.1 | 9.3 | C | B |
| C57 | - | -1.0 | 297.4 | 1.50 | 0.50 | 0.50 | 6.9 | 1.8 | 4.3 | C | A |
| C58 | - | -7.2 | 290.8 | 0.52 | 0.36 | 0.30 | 9.0 | 4.3 | 5.4 | C | B |
| C59 | - | -9.3 | 292.7 | 2.50 | 0.50 | 0.50 | 6.8 | 1.8 | 4.3 | C | A |
| C60 | - | -3.6 | 283.3 | 0.23 | 0.17 | 0.15 | 23.5 | 18.7 | 17.6 | C | B |
| C61 | - | -11.2 | 285.9 | 0.26 | 0.19 | 0.19 | 17.1 | 17.1 | 18.6 | C | B |
| C62 | - | -11.9 | 286.5 | 2.50 | 0.50 | 0.50 | 6.8 | 1.8 | 4.3 | C | A |
| C63 | - | -9.3 | 282.9 | 0.21 | 0.17 | 0.17 | 15.7 | 15.7 | 13.6 | C | B |
| C64 | - | -6.1 | 280.3 | 2.50 | 0.50 | 0.50 | 6.4 | 1.6 | 4.0 | C | A |
| C65 | M7 | -16.8 | 284.6 | 0.75 | 0.25 | 0.25 | 11.9 | 3.6 | 8.1 | S | A |
| C66 | - | -19.9 | 278.7 | 0.23 | 0.18 | 0.18 | 10.3 | 9.3 | 8.0 | C | B |
| C67 | - | -26.3 | 283.2 | 0.55 | 0.17 | 0.16 | 7.7 | 3.0 | 4.5 | C | B |
| C68 | - | -28.5 | 273.9 | 0.75 | 0.25 | 0.25 | 6.1 | 1.6 | 3.9 | C | A |
| C69 | - | 9.8 | 264.0 | 0.75 | 0.25 | 0.25 | 5.4 | 1.5 | 3.5 | C | A |
| C70 | - | -0.7 | 278.6 | 0.23 | 0.19 | 0.19 | 12.7 | 11.8 | 10.1 | C | B |
| C71 | - | 1.8 | 272.6 | 0.11 | 0.06 | 0.06 | 28.8 | 23.2 | 25.0 | C | B |
| C72 | - | -1.9 | 278.5 | 0.75 | 0.25 | 0.25 | 6.4 | 1.8 | 4.2 | C | A |
| C73 | - | -2.0 | 272.5 | 1.50 | 0.50 | 0.50 | 6.2 | 1.6 | 4.0 | C | A |
| C74 | - | -6.8 | 264.4 | 1.50 | 0.50 | 0.50 | 6.5 | 1.8 | 4.2 | C | A |
| C75 | - | 24.5 | 279.3 | 0.24 | 0.19 | 0.19 | 17.8 | 17.8 | 19.4 | C | B |
| C76 | - | -27.4 | 275.0 | 2.50 | 0.50 | 0.50 | 6.7 | 1.7 | 4.2 | C | A |
| C77 | - | 5.7 | 285.6 | 4.0 | 0.50 | 0.50 | 6.5 | 1.7 | 4.1 | C | A |
| C78 | - | -29.0 | 277.0 | 2.50 | 0.50 | 0.50 | 5.8 | 1.5 | 3.6 | C | A |
| C79 | - | -28.2 | 274.4 | 2.50 | 0.50 | 0.50 | 5.9 | 1.5 | 3.7 | C | A |

**Table 1.** (continued)

| ID | Latitude [deg] | Longitude [deg] | Diameter [m] | | | Normal Albedo [%] | | | Spectral Type§ [deg] | Size measurement Method* [deg] |
| --- | --- | --- | --- | --- | --- | --- | --- | --- | --- | --- |
| | | | $D_{ave}^{\#}$ | $\sigma_{D_l}^{\#}$ | $\sigma_{Du}^{\#}$ | $A_{normal}^{\dagger}$ | $\sigma_{A_l}^{\dagger}$ | $\sigma_{Au}^{\dagger}$ | | |
| S1 - | -3.4 | 295.6 | 0.22 | 0.21 | 0.15 | 16.41 | 6.09 | 7.73 | - | B |
| S2 - | 5.1 | 300.3 | 1.24 | 0.69 | 0.54 | 14.38 | 0.11 | 0.11 | - | A |
| S3 - | -1.0 | 299.0 | 2.79 | 1.79 | 1.35 | 14.12 | 0.61 | 0.61 | - | A |
| S4 - | 0.6 | 299.5 | 0.17 | 0.11 | 0.09 | 23.09 | 5.06 | 8.51 | - | B |
| S5 - | 0.7 | 299.7 | 0.60 | 0.28 | 0.22 | 10.99 | 0.35 | 0.35 | - | A |
| S6 - | 1.2 | 300.0 | 0.63 | 0.56 | 0.38 | 13.01 | 0.69 | 0.69 | - | A |
| S7 - | 1.2 | 301.1 | 0.26 | 0.08 | 0.07 | 22.38 | 0.65 | 4.38 | - | B |
| S8 - | 7.2 | 300.5 | 0.22 | 0.49 | 0.21 | 10.27 | 8.51 | 5.33 | - | B |
| S9 - | 7.9 | 301.6 | 0.39 | 0.59 | 0.35 | 9.59 | 4.62 | 4.31 | - | B |
| S10 - | 5.5 | 300.4 | 0.17 | 0.06 | 0.05 | 18.23 | 1.17 | 4.25 | - | B |
| S11 - | 6.0 | 300.0 | 0.11 | 0.05 | 0.04 | 13.19 | 1.27 | 3.06 | - | B |
| S12 - | 5.2 | 299.8 | 0.10 | 0.06 | 0.05 | 10.71 | 1.31 | 2.20 | - | B |

## 3. Size and spatial distributions of bright boulders

The new analysis conducted in this study revealed a number of important properties of bright boulders on Ryugu. In the following, we discuss the spatial distributions of bright boulders in Section 3.1, their SFD in Section 3.2, and their estimated volume and surface area in Section 3.3.

### 3.1 Spatial distribution of bright boulders

The coverage areas of the 2.7-km hovering observations (~0.29 m/pixel) and 1.7-km scanning observations (~0.18 m/pixels) are shown in Fig. 1. The observational coverage of the 2.7-km hovering images is limited to the equatorial region, but it includes all longitudes, which is useful for investigating the longitudinal distribution of bright boulders. In contrast, the 1.7-km scanning images cover higher latitudes, allowing us to examine the latitudinal distribution.

Bright boulders in the equatorial region exhibit a west/east dichotomy, which is similar to the longitudinal distribution seen in the general boulders (i.e., boulders with brightness 1.5 less than the median value of each image). Initial observations revealed that the western side of Ryugu (160 °E–290° E) has a lower number density of large boulders than the eastern side (Sugita et al., 2019; Michikami et al., 2019). In the images captured during the 2.7-km hovering observations, two to three bright boulders can be observed every 30° ranging between 240° E and 150° E (240° E to 0° E and 0° to 150° E). In contrast, no or only one bright boulder can be observed in the range between 150° E and 240° W. This observation suggests that bright boulders follow the same longitudinal distribution as the general boulders. The fact that both bright boulders and general boulders follow a similar longitudinal SFD suggests that the two types of boulders experienced similar size sorting processes, possibly as a result of the process leading to the west/east dichotomy formation.

The images captured during the 1.7-km scanning observations show that more bright boulders are present at lower latitudes than at higher latitudes. More specifically, 7, 35, 27, and 10 bright boulders are observed in the ranges of 90° N to 30° N, 30° N to 0° N, 0° S to 30° S, and 30° S to 60° S, respectively. Note that this apparent concentration of bright boulders in the equatorial region could be due to bias resulting from the different spatial resolution for different altitudes. The images of the high-latitude regions are distorted because of the greater tilt of the local surface with respect to the line sight of the ONC-T (Fig. 1). This leads to lower spatial resolutions at higher latitudes. It is also noted that the number density of large boulders (>10 cm) in general is much greater at higher latitudes than at lower latitudes (Sugita et al., 2019; Michikami et al., 2019). Thus, it is not clear that the latitudinal distribution observed for the bright boulders is different from that of large boulders in general. We underline that more extensive analysis is needed. Nevertheless, the fact that bright boulders are also observed at higher latitudes is evident from our results, indicating that bright boulders are ubiquitously present on Ryugu. This suggests that bright boulders are well mixed inside Ryugu's body, which might further suggest that they were mixed before Ryugu accreted to its current mass, as proposed by Tatsumi et al. (2021).

**3.2 SFD of bright boulders**

Regarding the SFD of bright boulders, two different sites on Ryugu, where high-resolution observations were conducted during the MINERVA-II deployment operation and TD1 rehearsal descents, show similar cumulative SFDs. The cumulative SFDs of bright boulders at 180°–265° E (TD1 rehearsal descents) are shown in Fig. 2a. These distributions express as:

$$N(> D) = N(> 1 \text{ m}) \left(\frac{D}{1 \text{ m}}\right)^{-\alpha}, \tag{3.1}$$

where $D$ is the diameter of the bright boulder and $\alpha$ is the power-law index. The best fit power-law indexes and their errors are listed in Table 2. Their power-law indexes range from 2.2 to 3.2. However, the power-law index over the entire size range is much steeper than the one estimated from narrower size ranges captured by individual images acquired at different altitudes. Comparisons between our results and the power-law indexes of bright boulders at different sites, 110°–140° E (Tatsumi et al., 2021) are shown in Fig. 3. The power-law indexes at these two sites are similar although some of them have large errors. The general agreement in power-law indexes between Tatsumi et al, (2021) and this study suggests that the SFD of bright boulders might be globally uniform.

We derived the best fit power-law indexes for the entire size range, using the following method. As described in Section 2.2, we defined objects brighter than 1.5 times the median value of each image as bright boulders. Due to the PSF, however, the albedo of small bright boulders that are only two pixels across or less are contaminated by light from the surrounding surface. Thus, they are not detected as bright boulders. Consequently, smaller bright boulders tend to be overlooked by our detection method. These small bright boulders need to be searched for in higher resolution images since this size of bright boulders suffer less from the "light dilution" due to the PSF effect in higher-resolution images. Thus, the SFD curves obtained with our method tend to be shallower (smaller $\alpha$) than the actual distribution. However, because the PSF is fixed on the CCD pixels, not on the actual physical size, the effect of the PSF on detection probability is expected to be the same for bright boulders covering the same number of pixels at different resolutions (i.e., m/pix). Thus, comparing the SFD at the same pixel diameter obtained in images with different spatial resolutions would provide a more accurate SFD slope (i.e., $\alpha$). Consequently, for the SFD derived from multiple images in Fig. 2a, the power-law indexes were calculated using data points with the same pixel size range (> 1 pixel). More specifically, we selected 6 data points (MINERVA-II deployment operation or TD1 rehearsal descents) from a relatively small size range and 24 data points (2.7-km hovering operation) from a relatively large size range and calculated the power-law indexes obtained with any combination (144 cases) of these data points as shown (Fig. 2b). Furthermore, regional differences in SFD are considered here. The SFDs derived from images captured during the 2.7-km hovering observation are integrated in Fig. 2a. In contrast, we used the SFDs obtained from the individual images in this analysis. Fig. 2b shows that there is a regional difference of about 10 times between images taken during 2.7-km hovering observation at different longitudes. The power-law indexes $\alpha$ were calculated by selecting 144 combinations from a relatively small size range and relatively large size range. The intercepts $N(> 1\text{ m})$ were calculated based on the obtained $\alpha$. The average index and intercept were $3.0 \pm 0.7$ and $11.4^{+4.0}_{-2.9}$, respectively (Fig. 2c).

The best fit power-law index for the entire size range from 0.02 to 2 m in diameter is approximately 3. Furthermore, it is noted that the power-law indices for the two different spectral types of bright boulders are very different. Using the same method described above, the best fit power-law index 1.6±1.3 for S-type bright boulders observed during the 2.7-km hovering observations was derived from 10 combinations of 5 data points (Fig. 2d). The large error for the power law index is resulted from the small number of data points. In fact, the distribution of number of cases is not close to Gaussian. However, we took $2\sigma$ as the error of the distribution, which encompasses the entire distribution obtained in the bootstrap calculation. Thus, it is likely that our error estimate is sufficiently conservative. The power-law index for C-type bright boulders is practically the same as that for all the bright boulders. This demonstrates that the power-law index for S-type bright boulders is much shallower than the C-type bright boulders value. Although the cause for such a difference is not fully understood, it may be due to the difference in the mechanical strength of C- and S-type bright boulders. The observation that S-type bright boulders have a shallower power-law index than C-type bright boulders is consistent with the finding that S-type material (i.e., OCs) generally has a much stronger mechanical strength than C-type material (i.e., CCs) (Flynn et al., 2018). Furthermore, the number density of C-type bright boulders is much greater than that of S-type bright boulders at the smaller size range. On the other hand, at the size range larger than ~1 m in diameter, the number density is dominated by S-type bright boulders.

Furthermore, lower thermal conductivities of CCs (e.g., Opeil et al., 2012) may contribute to fast breakdown through thermal fatigue of C-type bright boulders (e.g., Delbo et al., 2014). However, because the ranges of thermal conductivity among CCs and OCs are greater than the difference between their average values (Opeil et al., 2012), it is difficult to determine if this thermal conductivity difference between CCs and OCs contribute to the difference between S-type and C-type bright boulders until the actual thermal conductivity of Ryugu samples have been measured.

      Here, it is noted that some data points in Fig. 2a are multiple small bright clasts embedded in large boulders. Such counting is unavoidable because we automatically count regions with brightness 1.5 times the surroundings or greater. The entire boulders including multiple smaller bright clasts may be somewhat larger than the SFD shown in this figure. Thus, the SFD of bright boulders may be significantly different when the size of the entire bright boulders that contain multiple smaller bright clasts are used for the statistics. More specifically, the SFD or the power-law index for such a SFD of bright boulders that contain smaller bright clasts would have a shallower distribution than the distribution shown in Fig. 2a. Nevertheless, the general trend that S-type bright boulders are more abundant than C-type bright boulders at a larger size range will not change even after such a factor is considered. It is also important to note that the shallow SFD of S-type bright boulders has been confirmed by the higher-resolution, 1.7-km scanning observations as shown in Fig. 2a. We found one new S-type bright boulder in the 1.7-km scanning images used for the SFD analysis, whose size follows the shallow SFD defined by the S-type bright boulders in the 2.7 km-hovering observations (Fig. 2a). Thus, we decided to take this shallow power-law index ($1.6 \pm 1.3$) for the SFD of S-type bright boulders to estimate the volume and surface area of C- and S-type bright boulders in Section 3.3. Here, the error is $2\sigma$ (standard distribution) of the distribution.

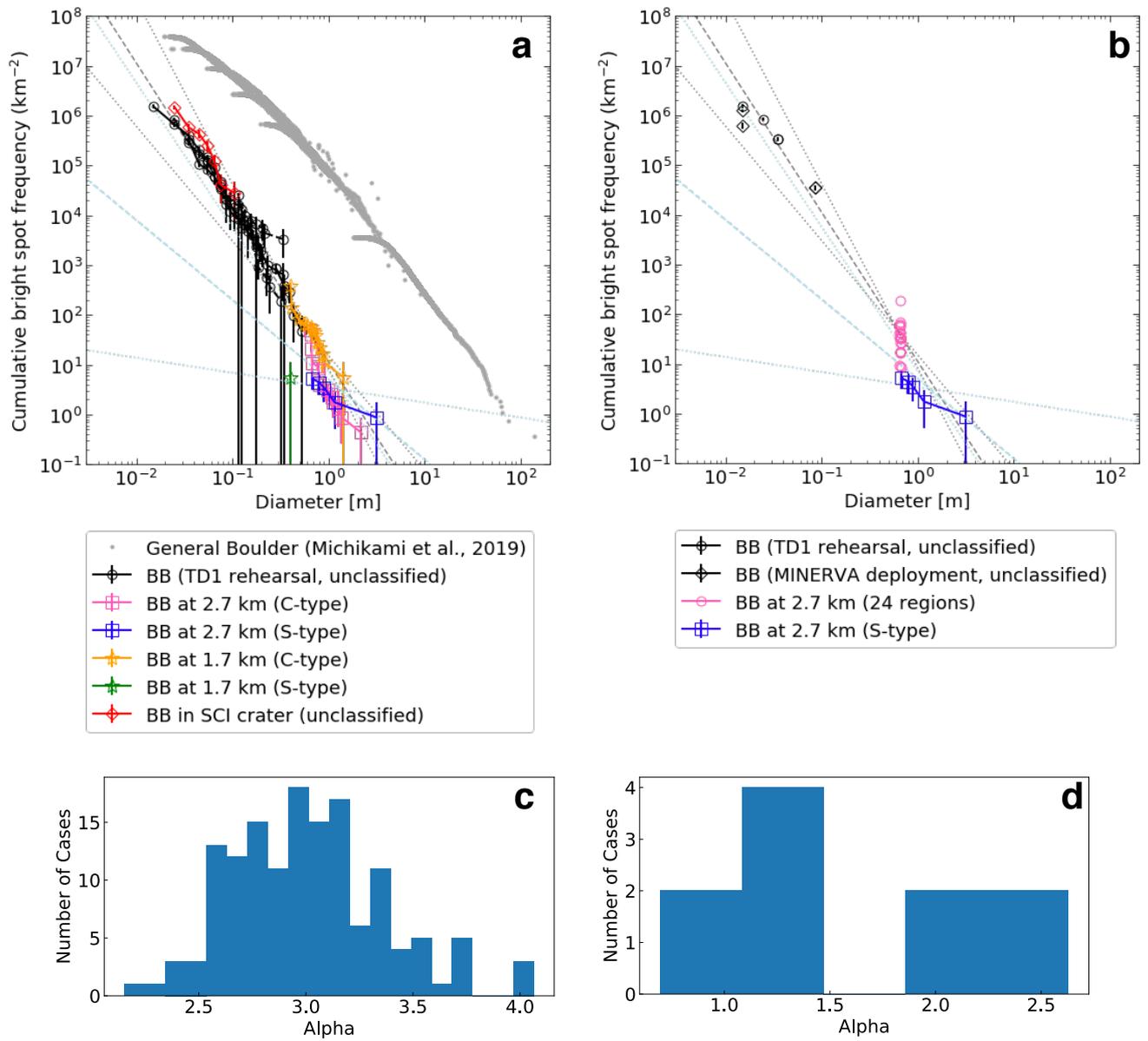

**Fig. 2** (a) The SFD of bright boulders and the bright boulders observed during a TD1 rehearsal (black circles), 2.7-km hovering observations (C-type: pink squares, S-type: blue squares), and 1.7-km scanning observations (C-type: yellow stars, S-type: green stars). Bright boulders in the SCI crater are indicated by red diamonds. General boulders studied by Michikami et al. (2019) are shown as grey circles for comparison. Dashed gray and blue lines indicate the best fit power-law distribution for C- and S-type bright boulders with power-law index 3.0 and 1.6, respectively (the power-law indexes and intercepts are listed in Table 3). Dotted gray and blue lines bound the upper and lower limits of these power-law indices 3.0±0.7 and 1.6±1.3. (b) SFDs used to obtain the best fit power-law indices for C/S-type bright boulders. Six data points (black diamonds:

MINERVA-II deployment operation, black circles: TD1 rehearsal descents) in the smaller size range and 24 data points (pink circles: 2.7-km hovering operations) in the larger size range were used. The best fit power-law index for C-type bright boulders was calculated from all the combination (144 cases) of these data points. The best fit power-law index of S-type bright boulders was derived from the SFD of S-type bright boulders observed during the 2.7-km hovering observation (blue squares). (c) The distribution of the power-law indices of C-type bright boulders calculated using the 144 combination of data points in Fig. 2b. (d) The distribution of the power-law indices of S-type bright boulders calculated using the 10 combination of data points in Fig. 2b.

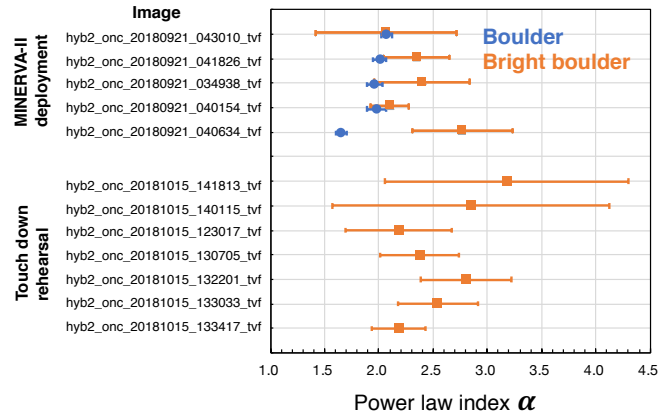

**Fig. 3** Best fit power-law index for boulder distributions using close-up images captured during MINERVA-II deployment and TD1 rehearsal operations. Power-law indices and their errors for bright boulders (orange squares) identified during MINERVA-II deployment are listed in Tatsumi et al. (2021) and those of bright boulders identified during TD1 rehearsal are listed in Table 2. Those of general boulders (blue circles) are listed in Michikami et al. (2019).

**Table 2** Power-law index for boulder distribution from close-up images captured during TD1 rehearsal observation.

| Image | Altitude (m) | Counted number of bright boulders | Power-law index | Error (1σ) of power index | Bright boulders size range (m) |
|---|---|---|---|---|---|
| hyb2_onc_20181015_141813_tvf_iof | 1194 | 18 | 3.18 | 1.12 | 0.29-0.53 |
| hyb2_onc_20181015_140115_tvf_iof | 617 | 41 | 2.85 | 1.27 | 0.16-0.32 |
| hyb2_onc_20181015_123017_tvf_iof | 464 | 78 | 2.19 | 0.49 | 0.1-0.35 |
| hyb2_onc_20181015_130705_tvf_iof | 241 | 135 | 2.38 | 0.37 | 0.07-0.34 |
| hyb2_onc_20181015_132201_tvf_iof | 159 | 141 | 2.80 | 0.42 | 0.05-0.18 |
| hyb2_onc_20181015_133033_tvf_iof | 100 | 154 | 2.55 | 0.37 | 0.03-0.12 |
| hyb2_onc_20181015_133417_tvf_iof | 80 | 182 | 2.18 | 0.25 | 0.02-0.12 |

### 3.3 Volume and surface area estimate for bright boulders

The SFDs of bright boulders observed on the surface of Ryugu or within the larger boulders on Ryugu were used to estimate the surface area and volumes of those materials. It is difficult to estimate how much more material is buried underneath other darker material on the surface or within the substrate boulders. In other words, we only consider bright boulders exposed on the surface. Using the SFD estimates for C-type and S-type bright boulders described in Section 3.2, we estimated both the volumes and surface areas of those two types of boulders.

Using the Eq. 3.1, the incremental SFD of bright boulders can be expressed by:

$$-\frac{\partial N(>D)}{\partial D} = N(>1\text{ m})\alpha\left(\frac{D}{1\text{ m}}\right)^{-\alpha-1}. \tag{3.2}$$

Assuming that boulders have the same aspect ratios as the dark boulders on Ryugu and other typical boulders on asteroids (Michikami et al., 2016; Michikami et al., 2019), the volumes and surface areas of bright boulders are given by:

$$\text{Volumes:} \int_{D_{min}}^{D_{max}} -\frac{\partial N(>D)}{\partial D}\frac{\pi}{6}D^3\,dD = \frac{\pi}{6}\frac{N(>1\text{ m})\,\alpha}{-\alpha+3}\left[D_{max}^{-\alpha+3} - D_{min}^{-\alpha+3}\right], \tag{3.3}$$

$$\text{Surface areas:} \int_{D_{min}}^{D_{max}} -\frac{\partial N(>D)}{\partial D}\frac{\pi}{4}D^2\,dD = \frac{\pi}{4}\frac{N(>1\text{ m})\,\alpha}{-\alpha+2}\left[D_{max}^{-\alpha+2} - D_{min}^{-\alpha+2}\right]. \tag{3.4}$$

The power-law indices $\alpha$, intercepts $N(>1\text{ m})$, minimum size $D_{min}$, and maximum size $D_{max}$ used to calculate the surface areas and volumes are listed in Table 3. Note that α=3.005 was used to avoid division by zero for C-type bright boulders. Here it is noted that actual boulders on Ryugu as well as other asteroids are known to have aspect ratios different from unity. However, the departure from a sphere is very small; the volumes of general boulders on Ryugu are estimated to be about 0.65 times of a sphere with the same dimension (Michikami and Hagermann, 2021).

The analysis results indicate that S-type bright boulders have a volume of about $3.7^{+3.3}_{-2.6} \times 10$ m³ and cover a surface area of $4.1^{+9.1}_{-3.2} \times 10$ m², while C-type bright boulders have a volume of about $2.3^{+7.3}_{-1.3} \times 10^2$ m³ and cover a surface area of $3.8^{+27.3}_{-3.1} \times 10^3$ m². The errors presented here were estimated in the following fashion. As discussed above in Section 3.2, we selected all the possible combinations of bright boulder abundance measurements at large diameters (~0.6 m) and those at small diameters (0.01–0.1 m). The fluctuation in boulder abundances among the different image data includes both statistical fluctuation and real spatial variation; we decided to consider both in our error estimation. Thus, our error estimates could be larger than the actual error. These volumes and surface areas are much smaller than the volume and surface area of the darker, general boulders.

The volume ratio of S-type bright boulders to darker boulders is only $7.1^{+6.3}_{-5.0} \times 10^{-6}$, while the surface ratio is $1.5^{+3.2}_{-1.2} \times 10^{-5}$. Note here that these ratios are effectively the ratio of S-type bright boulders to the entire Ryugu. Similarly, the ratios of C-type bright boulders to darker boulders are $4.4^{+14.0}_{-2.2} \times 10^{-5}$ in volume and $1.3^{+9.8}_{-1.1} \times 10^{-3}$ in surface area. For the diameter range 0.3 to 2 m overlapping between C-type and S-type bright boulders, the ratio of the total volume of C-type and S-type bright boulders to that of dark boulders are $1.8^{+0.3}_{-0.1} \times 10^{-5}$ and $3.9^{+6.9}_{-3.2} \times 10^{-6}$, respectively. The ratio of total surface area of C-type and S-type bright boulders to that of dark boulders at the same overlapping diameter range are $7.6^{+3.1}_{-1.9} \times 10^{-5}$ and $1.1^{+3.2}_{-0.9} \times 10^{-5}$, respectively.

If the hypothesis that S-type bright boulders on Ryugu are exogenic, as suggested by Tatsumi et al. (2021), is correct, then the mass mixing ratio of the projectile that hit the parent body of Ryugu is $7.1^{+6.3}_{-5.0} \times 10^{-6}$. This is important to constrain the nature of the collisional history of Ryugu's parent body. Furthermore, as shown by the small power-law index obtained above, the largest boulders account for the dominant proportion of the total mass of S-type bright boulders. In particular, the largest S-type bright boulder, M13, contributes significantly. More specifically, M13 has a semi-major and semi-minor axis of about $2.1 \pm 0.2$ m and $1.2 \pm 0.2$ m, respectively as shown in Fig. 4. Here, we used the average of the axes of the circumscribed and inscribed ellipses as the best fit value of the axes of this bright boulder and half the difference as the error.

Then, the volume of M13 is estimated to be $13 \pm 5 \text{ m}^3$, assuming that its two shorter axes are the same as the observed minor axis. This volume accounts for ~40% of the total volume of S-type bright boulders on Ryugu.

Similarly, the volume ratio $4.4^{+14.0}_{-2.2} \times 10^{-5}$ of C-type bright boulders is a suitable measure for the mixing ratio of the parent body's portions that may have experienced different thermal conditions from the dark general boulders on Ryugu. This small ratio of C-type bright boulders supports the theory that the vast majority of Ryugu's materials experienced uniform degrees of thermal metamorphism on Ryugu's parent body. This also places import constraints on the evolution of Ryugu's parent body. More specifically, such a uniform thermal history over a large portion of Ryugu's parent body supports uniform heating processes such as radiogenic heating in Ryugu's parent body as proposed by Sugita et al. (2019).

Furthermore, the fact that the surface area ratio of C-type bright boulders is quite large ($1.3^{+9.8}_{-1.1} \times 10^{-3}$) also has an important implication for understanding the very high-resolution images of Ryugu and possibly the returned samples from Ryugu to Earth. This ratio is calculated based on the size range down to only 2 cm. The SFD of bright boulders smaller than 5 cm has a rollover in the obtained data. However, this may be due to the insufficient spatial resolution of the images. Thus, the same steep SFD might extend to smaller size ranges for bright boulders, possibly down to a mm in diameter. If this was the case, then the ratio of C-type bright boulders could be even higher. For example, using the average estimated power-law index (3.0) for C-type bright boulders, the power-law index (1.65) for general boulders with sizes between 0.02–3.26 m (Michikami et al., 2019), and the size range 1 mm to 1 cm (the approximate size of the sampler horn), the surface area ratio of C-type bright boulders is ~30%. Due to the error in the estimated power-law index of C-type bright boulders, this volume ratio estimation has a large error. Nevertheless, when we use the lower limit of the estimated power-law index (2.3) for C-type bright boulders, the surface area ratio of C-type bright boulders is estimated to be as much as ≳0.4%. Thus, the odds that C-type bright boulder materials are contained in the samples captured by Hayabusa2 touchdown could be significant. In particular, the finding that some of the boulders observed on the Ryugu surface by MASCOT have large bright inclusions with the size range of sub-millimeter to a few millimeters is not inconsistent with the SFD extrapolated from the C-type bright boulders in this study (Jaumann et al., 2019). The nature of such bright inclusions on dark boulders on Ryugu's surface may be a very important observational constraint on the nature of Ryugu's surface material. Thus, the search and examination of such potential inclusions or C-type bright materials will be of very high interest in the returned samples.

**Table 3** Power-law index, intercept and size range used to calculate the surface area and volume of C/S-type bright boulders.

| | Power-law index $\alpha \pm 2\sigma_{alpha}$ | Intercepts $N(> 1\ m)$† | Bright boulders size range (m) |
|---|---|---|---|
| C-type bright boulder | $3.0 \pm 0.7$ | $11.4^{+4.0}_{-2.9}$ | 0.02-2 |
| S-type bright boulder | $1.6 \pm 1.3$ | $5.0^{+2.3}_{-1.6}$ | 0.3-3 |

†: The upper and lower errors are from intercept N(>1 m) for steepest (α+2σ) and shallowest (α–2σ) SFDs. Respective combinations of intercept and power-law index are used, the upper and lower estimates for volume and surface area of bright boulders are obtained.

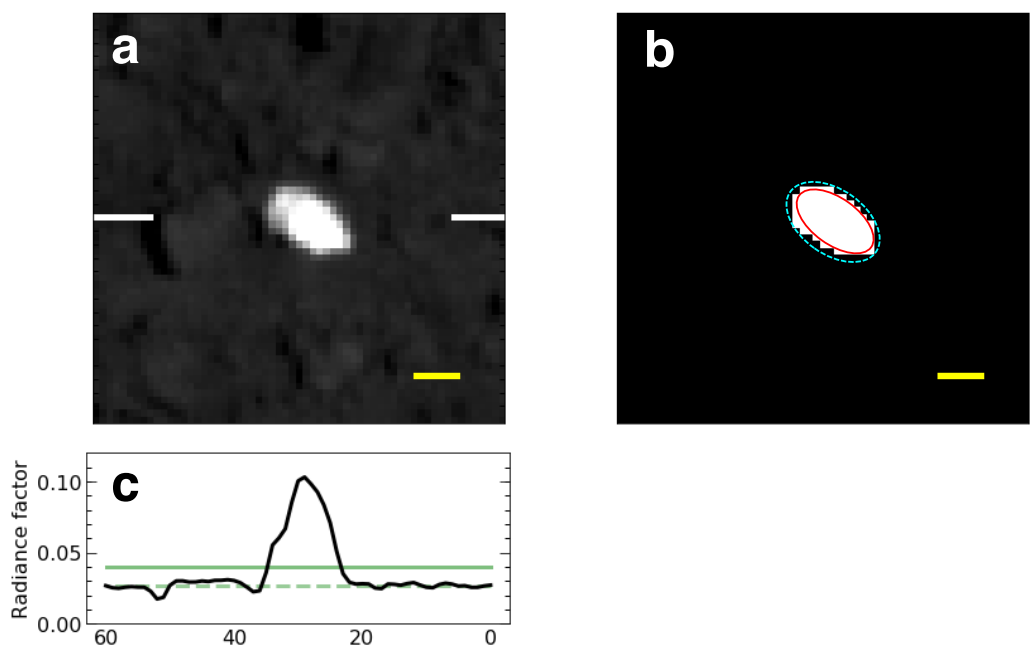

**Fig. 4** (a)V-band image (hyb2_onc_20181003_232509) of an S-type bright boulder (M13) found by Tatsumi et al. (2021). The yellow scale bars is 2 m. (b) The semi-major and semi-minor axes of the circumscribed ellipse (cyan dashed line) of the bright boulder, whose area is defined by brightness values greater than 1.5 times the median value of the background, are 2.3 and 1.4 m, respectively, and those of the inscribed ellipse (red solid line) are 1.9 and 1.0 m, respectively. (c) Radiance factor profiles along the white lines in (a). Green dashed line and solid line indicate the median value of the background and 1.5 times of it, respectively.

## 4. Occurrence and morphologies of bright boulders

In this section we discuss the morphologies of bright boulders mostly based on higher resolution (down to ~1 cm/pixel) observations. Morphological analysis based on high-resolution images captured at low altitudes revealed two notable characteristics of the bright boulders on Ryugu. 1) Some bright boulders have intra-boulder variation in reflectance and 2) bright boulders are also observed inside the artificial crater cavity newly formed by the SCI. These findings have important implications for the origin of bright boulders.

Some of the spatially resolved bright boulders are observed as small individual boulders (Figs. 5a and b). Reflectance profiles across these bright boulders are shown in Figs. 5c and d. The reflectance profile shown in Fig. 5c shows that the jump in reflectance at the boundaries of the boulder is much greater than the background fluctuation in reflectance on Ryugu. Specifically, the reflectance of this bright boulder is ~1.5 times brighter than that of the surrounding background and the reflectance jump between the plateau of the bright boulder and background is >4 $\sigma_{bk}$ where $\sigma_{bk}$ is the standard deviation in the background reflectance in the image. The reflectance profile shown in Fig. 5d shows a much greater jump (~60$\sigma_{bk}$) at the boundaries of the bright boulder and shows that this bright boulder is ~3.7 times brighter than the surrounding background. These bright boulders have C- and S-type spectra, respectively as shown in Fig. 5e.

Other bright boulders are observed as bright clasts embedded in larger substrates (Figs. 6a, b, c, and possibly d). The presence of such bright clasts embedded in general boulders is important because such morphologies indicate that these bright clasts have experienced intense mixing and agglomeration within darker fragments in Ryugu's parent body rather than having soft-landed on Ryugu in its current small size. Reflectance profiles across these bright clasts are shown in the same figures along with the boundaries of the bright area and substrate boulder. The increases in the reflectance of these intra-boulder bright clasts are also much larger (e.g., ~20$\sigma_{bk}$ in Fig. 6a) than the background fluctuation $\sigma_{bk}$.

Furthermore, multiple bright spots are sometimes seen in a single substrate boulder (Figs. 6a, b, c, and e). The appearances observed so far are all consistent with bright clasts embedded in a large substrate boulder. For example, one boulder has two intra-boulder bright clasts (Fig. 6a). No apparent shadows are observed around these intra-boulder bright clasts, whereas the boundary at the base of the boulder that contains these bright spots displays a clear shadow. Meanwhile, other boulders of comparable sizes in this image also display shadows. Similarly, in Fig. 6b, another bright spot has a small extension to the right, and it does not cast a shadow on the substrate boulder. If the bright spot is topographically higher than the dark substrate boulder, shadows would be observed. Furthermore, the shadows of this bright spot and substrate boulder in the upper left are smoothly connected; there are no abrupt kinks in the shadow boundary. This observation suggests that they have a continuous surface. A third bright spot is found to display a two-step increases in reflectance (Fig. 6c). Most of the area for this bright spot is approximately two times brighter than the surrounding background, but a small brighter cluster (at least approximately four times brighter than the background) is included within this area. No apparent shadows are observed around the larger bright spot or the smaller bright cluster. The left side of this boulder is darker than the surrounding terrain. However, this dark region is not a shadow but rather a region with a high solar incidence angle due to a large and sharp change in surface inclination. In contrast, the right side of this boulder has a reflectance comparable to the background; the upper and lower edges of the bright spot and the right side of the boulder are smoothly connected. This smooth connection suggests that their surfaces are continuous. Thus, the significant brightness variations on this boulder are likely due to the variations in the constituent material. We also found additional bright spots, whose morphologies support embedded structures, but not as strongly as the above cases. One such example is shown in Fig. 6d. On this boulder, the substrate boulder exhibits imbricated structures; shadows are projected to the left of a convex dark mass on this boulder. The right side of the bright boulder appears to be partially hidden by the shadow of a boulder sitting on the substrate boulder. This observation suggests that this bright spot is flat, which is consistent with an embedded structure for this bright spot.

In the four images described above, reflectance profiles suggest that the background brightness is ~1/3 of the bright spots but the CCD pixels of the brightest area are saturated. Thus, the brightest points of these boulders have a reflectance of at least ~3 times that of Ryugu's surface. As described in Section 3 in Paper 2, S-type bright boulders tend to be brighter than C-types. Among relatively well-resolved bright boulders ($\geqq$ 10 pixels) studied by Tatsumi et al. (2021) and in this study, all bright boulders with reflectance >3 times the global average of Ryugu exhibit S-types spectra. The reflectance factors of C-type bright boulders range from 1.1 to 2.3 times the global average, while S-types are 1.8–4.7 times. Thus, bright clasts with such high reflectance are more likely S-types than C-types. This observation suggests that many S-type fragments, which are very likely exogenic, may be embedded in Ryugu's boulders.

To investigate such a possibility, we examined the spectra of bright clasts embedded in dark boulders. However, the images with multiple color band observations of bright clasts are not as high in spatial resolution as the ones used for the morphological analysis presented above. The multi-color observations of bright boulders within the substrate boulders are available in ONC dataset; shown in Figs. 6f, g, and h (bright boulders C6, S2, and C65 in Table 1, respectively). Comparisons between substrate boulder and the embedded clasts indicates that these bright clasts possess color and reflectance properties different from those of the substrates (Fig. 7). The substrates in Figs. 6f, g, and h exhibit spectra similar to the global average of Ryugu. In contrast, as described in Section 3 in Paper 2, the bright clasts, such as C65 and C6, on these substrate boulders exhibit S-type or S-type candidate spectra, respectively (Fig. 7). Both bright clasts and the substrate boulder in Fig. 6g exhibit relatively flat ul-to-v slope, which is defined by the slope between ul and v bands on ONC-T (Tatsumi et al., 2021), but the bright spot has a slightly stronger UV-downturn.

One of the S-type bright boulders is also observed as a bright spot on a larger darker boulder (Fig. 6h). A super-resolution image composed using the same method by Lucy (1974) and Richardson (1972) based on 4 individual images of this boulder is shown in Fig. 8. In this image, the shadow of the S-type bright boulder can be observed at the west part of the boulder, suggesting that a part of the bright boulder is extended beyond the edge of the substrate boulder. The ratio of the area extending beyond the edge to that within the substrate boulder is ~1.5. This unstable configuration indicates the presence of some kind of cohesion between the bright boulder and substrate boulder. This observation strongly suggests that the substrate boulder is a polymict breccia that includes this S-type clast. This is consistent with the hypothesis that S-type bright boulders are more likely to be mixed during and/or before the catastrophic disruption of Ryugu's parent body and subsequently cemented together with other Ryugu materials rather than to have soft-landed on Ryugu after its formation, as suggested by Tatsumi et al. (2021).

The SCI experiment performed by the Hayabusa2 mission provides important information on the occurrence and distribution of bright boulders. Hayabusa2 performed the first detailed examination of an artificial impact crater (i.e., SCI crater), especially on a rubble-pile asteroid, created by the SCI on April 5, 2019 (Arakawa et al., 2020). Bright boulders are observed around and inside the SCI crater (Fig. 9). The disturbance of the surface due to this artificial impact provided a unique opportunity to examine whether bright boulders are mechanically adhered to substrate boulders and whether bright boulders exist underneath the surface layers of Ryugu.

Image comparisons between before and after the SCI experiments indicate that a bright boulder (S2) was not displaced from its substrate boulder through the SCI impact, as shown in Fig. 9. Note that S1–S12 bright boulders in this figure are selected by visual inspection. The observation suggests that this bright boulder is embedded in the substrate boulder. More specifically, the substrate boulder (labeled Stable Boulder (SB) in Fig. 9) is located in the vicinity (~3 m) of the impact point and exhibited no appreciable motion (<1 pix; 1.3 m) caused by the SCI impact (Arakawa et al., 2020). However, other similar sized boulders, such as the one labeled Mobile Boulder (MB), were significantly moved by the impact and a large volume of material was excavated to the north and west of the stable boulder (Arakawa et al., 2020). Therefore, the Stable Boulder must have experienced intense disturbance by the impact. Nevertheless, this bright boulder on Stable Boulder was not moved by the disturbance, suggesting that this bright boulder is mechanically adhered to Stable Boulder.

Furthermore, in the high-resolution post-SCI images, bright boulders can be observed on the floor of the artificial SCI crater, suggesting that bright boulders also exist in the subsurface layers of Ryugu (Figs. 9d and e). Observations of such subsurface bright boulders are important for at least a couple of different reasons. First, both the crater morphology observations (Cho et al., 2020) and the SCI ejecta curtain observations (Kadono et al., 2020) suggest that there may be subsurface stratifications in the boulder SFD, indicating the possibility of size sorting. Such size sorting may influence the spatial distribution of bright boulders on Ryugu. Boulder size stratification may also be important for understanding the difference between surface mass motion on Ryugu and Bennu. Bennu exhibits more localized mass motions (Walsh et al. 2019; Jawin et al., 2020) than Ryugu, which exhibits region-scale (e.g., equatorial ridge toward mid-latitude geopotential lows) mass motions (Sugita et al., 2019; Morota et al., 2020). Second, it is highly unlikely that bright boulders newly excavated by the SCI experiment experienced space weathering. Thus, their spectral and albedo properties are important for understanding the intrinsic compositional properties of bright boulders on Ryugu.

However, not many spatially resolved bright boulders have been found on the SCI crater floor owing to its limited surface area. Furthermore, because some filter-band images are saturated by the albedo of these bright boulders, their exact spectral properties cannot be determined with great confidence. Nevertheless, some v-band images were not saturated by these bright boulders (such as S10, S11, and S12 in Fig. 9). Thus, we were able to estimate the approximate brightness of these bright boulders. Analysis results indicate that bright boulder S10 exhibits a brightness 2.7 times the brightness of its surroundings. Because this brightness contrast is significantly greater than the other C-type bright boulders (≲2.3 times) on Ryugu, this may have S-type spectrum or other bright substance, such as carbonates as found on Bennu (Kaplan et al., 2020). Bright boulders S11 and S12 exhibit 1.9 times the brightness of its surroundings, which are consistent with the brightness of both C- and S-type bright boulders observed thus far.

Furthermore, using the same method described in Section 2.2, we studied the SFD of bright boulders within the SCI crater floor (Fig. 2). The SFD of the SCI crater floor bright boulders follow the entire SFD of bright boulders measured in various observations, which are previously discussed in Section 3 (the TD1 rehearsal, the 2.7-km hovering observations, and the 1.7-km scanning observations). Because Arakawa et al. (2020) and Kadono et al. (2021) show that most of the material within the SCI cavity experienced only weak shock pressures (≲ 500 Pa), fragmentation of bright boulders due to the SCI impact experiment is expected to be minimal. Thus, this observation suggests that the number density of bright boulders is not greatly influenced by the possible size-sorting processes on Ryugu.

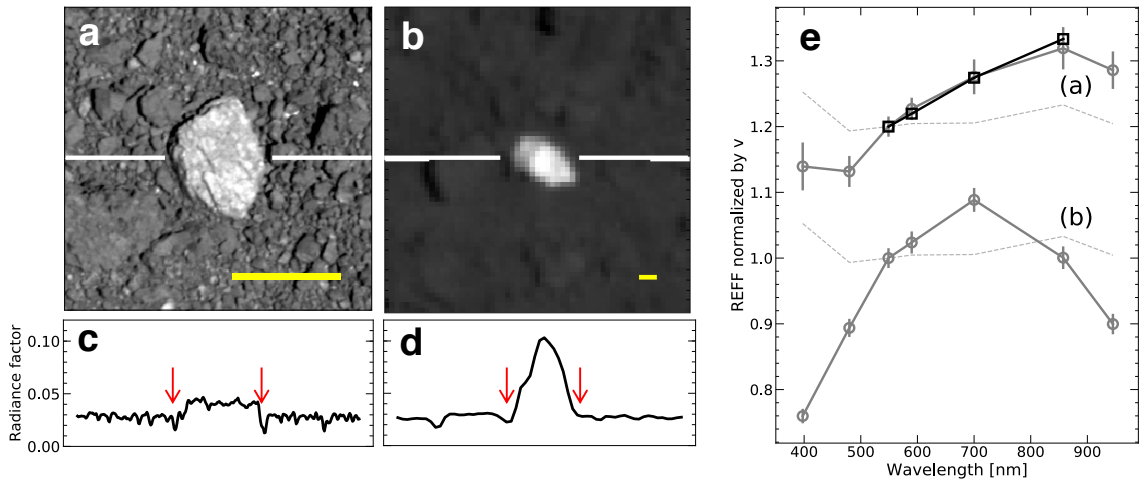

**Fig. 5** Individual bright boulders on Ryugu. The yellow scale bars correspond to 1 m. (a) A v-band image (hyb2_onc_20190221_215938) of a C-type bright boulder. This is the same boulder with an anomalously high thermal inertia shown in Fig. 4 from Okada et al. (2020). (b) A v-band image (hyb2_onc_20181003_232509) of an S-type bright boulder (M13) found by Tatsumi et al. (2021). This is the same image as Fig. 4, but its brightness level has been stretched to make the intra-boulder structure more discernible. (c) (d) Radiance factor profiles along the white lines in (a) and (b), respectively. Red arrows indicate the boundaries of the boulder. (e) The v-band normalized reflectance factor (REFF) spectra of bright boulders in (a) and (b). Gray lines with circles are the spectra from the 7-band images captured during the MASCOT hovering operation at ~2.7 km altitude on Oct. 3 – 4, 2018. The black line with squares is the spectrum from the 4-band images captured at an altitude of ~100 m on Feb. 21, 2019. Note that the latter spectrum is not photometrically corrected. Due to Ryugu's rotation, this bright boulder was observed in only 4 band images on Feb. 21, 2019. Images taken in the other three bands did not capture this bright boulder in the FOV (details of the spectral analysis are provided in Paper 2).

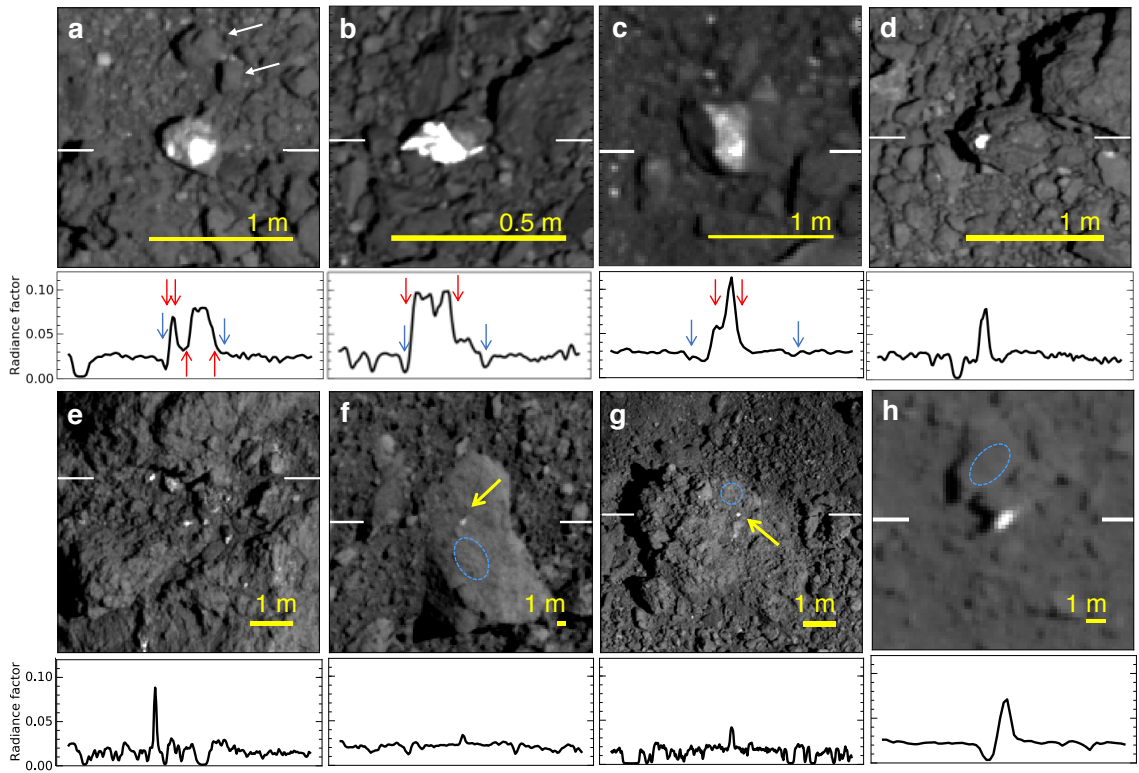

**Fig. 6** Intra-boulder bright clasts. The v-band image of each bright boulder is shown in the upper panel, and the radiance factor profiles are shown in the lower panels similar to that in Fig. 5. The boundaries of the bright parts and substrate boulders are indicated with red and blue arrows in (a)–(c). The digital numbers of the brightest area in (a)–(d) are saturated. Thus, these bright clasts are brighter than those shown in the radiance factor profiles. (a) hyb2_onc_20190308_030548; white arrows in panel (a) indicate dark boulders with sizes comparable to that of the bright boulders in this image. (b) hyb2_onc_20190308_032448, (c) hyb2_onc_20181015_130705, (d) hyb2_ onc_20190221_215938, (e) hyb2_onc_20190613_010043, (f) C6 bright boulder; hyb2_onc_ 20190321_190554, (g) hyb2_onc_20190613_020404, and (h) C65 (M7), hyb2_onc_20190321_193410. The blue dashed ellipses shown in (f), (g), and (h) are the areas used for the spectral measurements shown in Fig. 7.

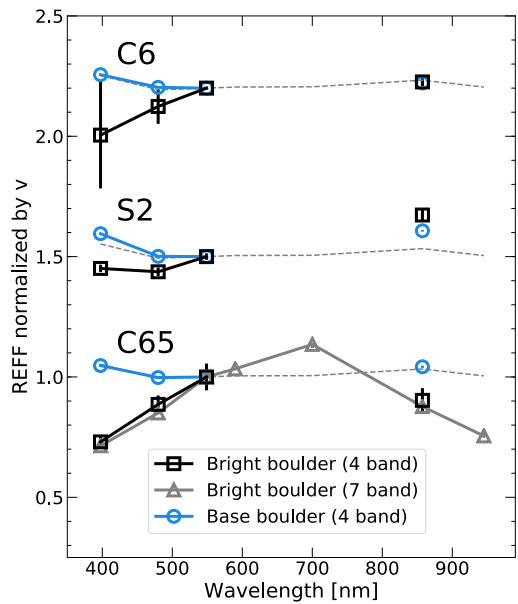

**Fig. 7** Spectral comparison between intra-boulder bright clasts and their substrate boulders shown in Fig. 6. Black lines with squares are the spectra of the bright clasts and the blue lines with circles are the spectra of the substrate boulders: the spectra of bright boulders/spots C6, S2, and C65. Dashed gray lines are the global average spectrum of Ryugu. The gray line with triangles is the spectrum of the same boulder derived from the 7-band images captured during the 2.7-km hovering observations (details of the spectral analysis are provided in Paper 2).

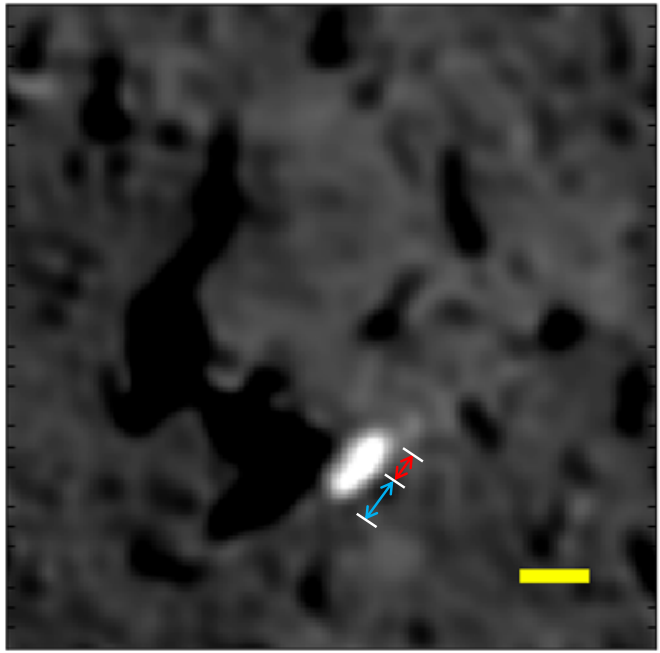

**Fig. 8** Super-resolution image of C65 (M7) bright boulder. The yellow scale bar is 1 m. The red arrow indicate the part of the bright boulder that appears to be attached to the substrate boulder, while the blue arrow indicates that this part of bright boulder extends out of the substrate boulder. The ratio of the areas indicated with the red and green dashed lines is ~2:3.

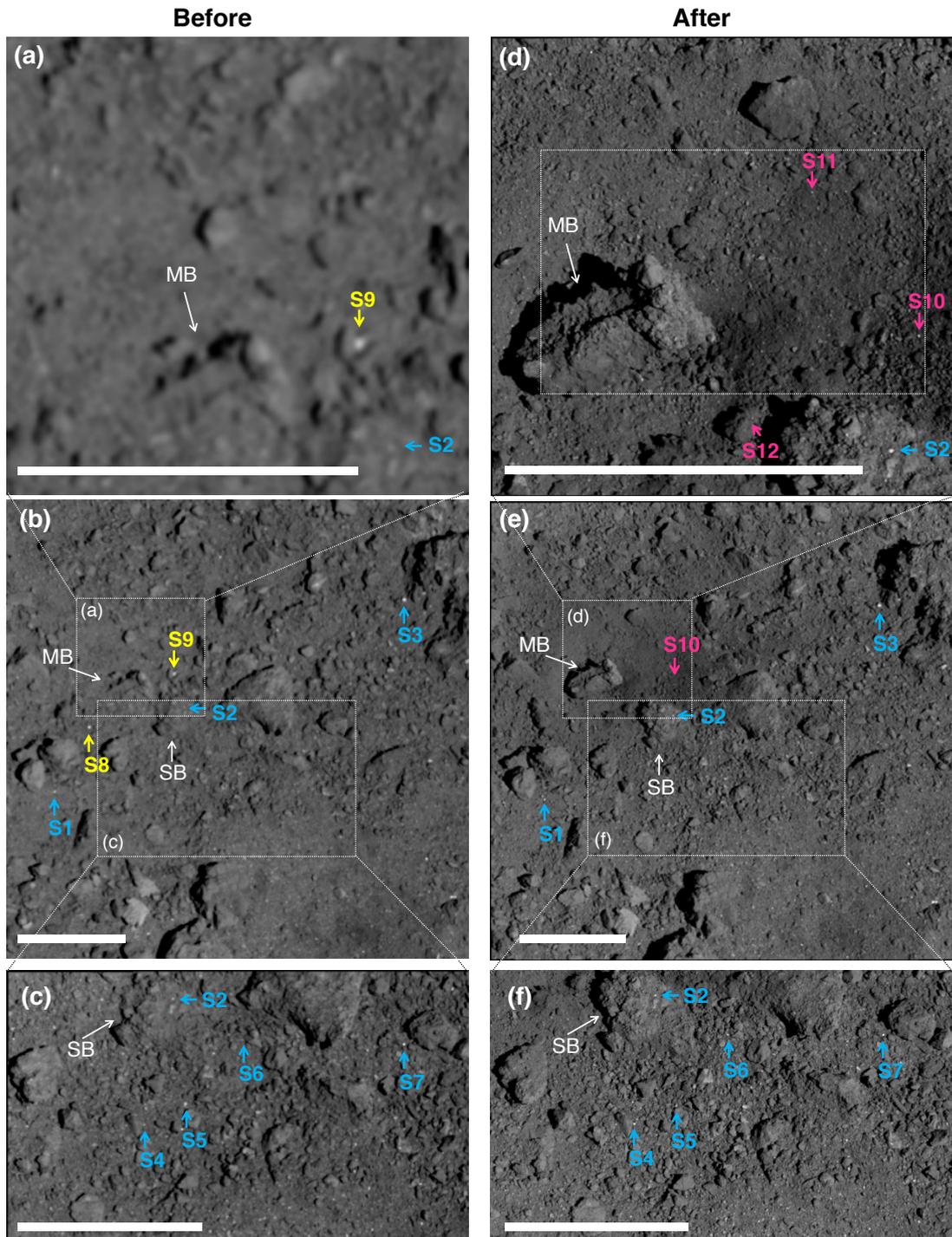

**Fig. 9** Comparison of bright boulders around the SCI crater before and after the impact. Size bars are 1 m. The left (a-c) and right (d-f) column images were captured before and after the SCI impact experiment, respectively. The upper and bottom images are close-up images of the center images. The Mobile Block (MB) and Stable Block (SB) are indicated with white arrows. Large boulders and bright boulders #1–#7 with cyan arrows in the south of the SCI crater remained at the same location before and after the impact experiment. In contrast, large and bright boulders #8–#9 with yellow arrows inside the SCI crater were moved or driven away from the surface. Bright boulders #10–#14 with magenta arrows are newly observed in the floor of the SCI crater cavity. Note that some of the bright boulders, such as S4 and S6, are faint in a pre-impact image (c) and became

much clearer in a post-impact image (f); (a, b) hyb2_onc_20190308_035656, (c) hyb2_onc_20190308_033742, (d) hyb2_onc_2019 0613_020224, (e) hyb2_onc_ 20190613_021711, and (f) hyb2_onc_ 20190613_020918.

# 5. Implications for collisional history and brecciation on Ryugu's parent body

The results from our morphological analyses of bright spots (boulders and clasts) also have important implications for Ryugu's collisional history. First, as discussed in Section 4, some bright spots appear as bright clasts embedded within larger substrate boulders. In particular, one of the S-type bright clasts exhibits a morphology indicating mechanical adherence to the substrate boulder. Such morphologies indicate that they have experienced mixing and agglomeration with darker fragments, which constitute the dominant fraction of the host rocks on Ryugu's parent body, rather than having soft-landed on Ryugu after its re-accretion. Thus, the presence of structurally embedded S-type bright boulders supports the hypothesis that these boulders were likely mixed during/before the catastrophic disruption of the parent body that created Ryugu (Tatsumi et al., 2021).

Second, the results from our detailed analyses of bright clasts embedded in larger substrate boulders can constrain the timing and process of brecciation on Ryugu's parent body. Furthermore, S-type and C-type intra-boulder bright clasts have different types of constraints. For example, one of the intra-boulder bright clasts exhibits an S-type spectrum and lacks a $0.7-\mu$m absorption. This observation indicates that the mafic minerals, which are rich in S-type minerals, have not been aqueously altered to serpentine. Thus, the delivery of this S-type intra-boulder bright clast material, as well as its brecciation or cementation in the surrounding dark C-type clasts, likely occurred after the cessation of aqueous alteration on Ryugu's parent body. The other intra-boulder bright clasts exhibit C-type spectra. As described in Section 4.2 in Paper 2, C-type bright clasts may have experienced different degrees of thermal metamorphism from dark general boulder materials on the parent body. Thus, cementation processes that combine bright C-type clasts and general dark boulder materials need to occur after the cessation of thermal metamorphism in the original parent body of Ryugu.

These results indicate that the cementation process on Ryugu's parent body may have continued even after the cessation of aqueous alteration and possible thermal metamorphism and may have taken place after the catastrophic disruption of the parent body (i.e., Polana or Eulalia), which is at least a billion years after the parent body formation (Bottke et al., 2015). Thus, the embedding of an S-type fragment in a large breccia on Ryugu suggests that the cementation process might have occurred billions of years after the formation of the Solar System.

If the cementation on Ryugu's parent body occurred over such a long period or over many stages of its evolution, a large number of impact fragments generated over the long history of Ryugu's parent body could cement together to generate many breccias. These could be the dominant constituent materials on Ryugu's parent body. The dominance of brecciated material may contribute to the globally low thermal inertia of Ryugu (Sugita et al., 2019; Grott et al., 2019; Okada et al., 2020; Shimaki et al., 2020). Indeed, it is known that an increase in particle size from regolith to boulders typically leads to increased thermal inertia. However, the spatial distribution of thermal inertia on Ryugu does not appear to follow this typical trend; regolith and boulders exhibit practically the same thermal inertia on Ryugu. Similarly, a more unusual trend is observed on Bennu; the thermal inertias of boulders are lower than its regolith (DellaGiustina et al., 2019; Ryan et al., 2019).

Furthermore, the dominance of brecciated materials is further consistent with the properties of carbonaceous chondrites. Large fractions of dark carbonaceous chondrites (CM and CI chondrites) are known to be breccias; ~100% of CM and CI meteorites are breccias (Bischoff et al., 2006). Clasts from some chondrites show various degrees of aqueous alteration (e.g., Kerraouch et al., 2019). The late brecciation on Ryugu's parent body, however, raises a question regarding what process caused the cementation of clasts within breccias on Ryugu. Although the breccias within carbonaceous chondrites may form when aqueous alteration occurs in carbonaceous chondrite parent bodies (e.g., McCoy et al., 2019), the above interpretation that brecciation occurred after the cessation of aqueous alteration on Ryugu is inconsistent with cementation by aqueous alteration processes. Another possibility is cementation via impact melt (e.g., Bischoff et al., 2006). Although impact melts would have played an important role in breccia formation, creating a sufficient amount of impact melt to cause adherence among a large fraction of clastic materials may not be easy to generate. This is because the volume of materials that experiences sufficiently intense shock heating to reach the melting temperatures of silicates (e.g., ~1000 °C) is much smaller than the mass of cold fragments from the catastrophic disruption of a 100-km parent body (e.g., Michel and Ballouz et al., 2020; Jutzi and Michel 2020). However, precisely quantifying the volume fraction of melting that is necessary or which phase of materials is needed to melt for abundant breccia formation remains a challenge. The key to answering this question may reside in the properties of the adhesive or cementing materials that glue together clasts within the breccia of carbonaceous chondrites. However, cementing materials in CC breccias are not well understood because of the unavoidable oxidation (i.e., rusting) of meteoritic materials after their fall to Earth. Thus, samples of Ryugu brought back by Hayabusa2 and those of Bennu returned by OSIRIS-Rex without exposing the samples to either rain or oxygen-rich air, will be suitable samples to investigate these adhesive materials. When the nature of this material that cements breccias on Ryugu is understood, then we will gain insight into the brecciation process on Ryugu's parent body.

# 6. Conclusions

We analyzed 12 high-resolution (~0.19 m/pixel), 0.55-$\mu$m images obtained during the 1.7-km scanning observations and 11 higher-resolution (down to 3.8 mm/pix) images obtained during the low-altitude descents (down to ~35 m). The analysis of more than a thousand bright spots (brightness ≥1.5 times the surrounding regolith and boulders) detected in our measurements yielded the following conclusions:

- Bright boulders at higher latitudes (>30° S and >15° N) were found, indicating that bright boulders are present ubiquitously on Ryugu. This ubiquitous presence suggests that bright boulders are well mixed in the body of Ryugu, supporting the hypothesis that they are mixed before rather than after Ryugu's accretion.
- Bright boulders in the equatorial region exhibit a west/east dichotomy in abundance. The longitudinal pattern is the same as the dichotomic pattern found for large dark boulders by Sugita et al. (2019) and Michikami et al. (2019).
- The SFD of S-type bright boulders in the range 0.3 m to 3 m follows a power law with an exponent of 1.6±1.3.
- Comparisons in SFD between S-type bright boulders and darker general boulders indicate that that the former has a volume of about $3.7^{+3.3}_{-2.6} \times 10$ m$^3$ and surface area of $4.1^{+9.1}_{-3.2} \times 10$ m$^2$. The volume ratio of the former to the latter is only $7.1^{+6.3}_{-5.0} \times 10^{-6}$ and the surface ratio is $1.5^{+3.2}_{-1.2} \times 10^{-5}$.
- The SFD of C-type bright boulders from 2 cm to 2 m follows a power law with an exponent of 3.0±0.7.
- Comparisons in SFDs between C-type bright boulders and general boulders indicate that that the former (diameter from 2 cm to 2 m) has a volume of about $2.3^{+7.3}_{-1.3} \times 10^2$ m³ and a surface area of $3.8^{+27.3}_{-3.1} \times 10^3$ m². The volume ratio of the former to the latter is $4.4^{+14.0}_{-2.2} \times 10^{-5}$, and the surface ratio is $1.3^{+9.8}_{-1.1} \times 10^{-3}$. When this power law is extrapolated to millimeter scales, which is the size range of samples that can be captured by the Hayabusa2 sampler system, the abundance of C-type bright boulder materials is estimated to be greater than 0.4%.

- For the diameter range 0.3 m to 2 m overlapping between C-type and S-type bright boulders, the ratio of the total volume of C-type and S-type bright boulders to that of dark boulders are $1.8^{+0.3}_{-0.1} \times 10^{-5}$ and $3.9^{+6.9}_{-3.2} \times 10^{-6}$, respectively. The ratio of total surface area of C-type and S-type bright boulders to that of dark boulders at the same overlapping diameter range are $7.6^{+3.1}_{-1.9} \times 10^{-5}$ and $1.1^{+3.2}_{-0.9} \times 10^{-5}$, respectively.

- The number density of bright boulders inside the newly made SCI crater is consistent with the outside number density to within a factor of two. This confirms the assumption that the abundance of bright boulders is similar between the surface and subsurface; the observed bright boulder abundance represents the bulk abundance of bright boulders on Ryugu.

- Many bright clasts embedded in large substrate boulders were observed. These bright clasts include both S-type and C-type materials. Such morphologies indicate that they have experienced mixing and agglomeration with darker fragments on Ryugu's parent body rather than a soft-landing on Ryugu after Ryugu's accretion. This observation also suggests that the cementation process on Ryugu's parent body may have continued until after the cessation of aqueous alteration, thermal metamorphism, and perhaps until catastrophic disruption of the parent body, resulting in a very large number of breccias. Such breccias would have a very high porosity and may have contributed to the globally low thermal inertia observed for Ryugu.

**Acknowledgments.** The authors wish to thank all members of the JAXA's Hayabusa2 team for their operational support and scientific discussions. Reflectance spectra of meteorite samples used in this study are from RELAB database. This study was supported by the Japan Society for the Promotion of Science (JSPS) Core-to-Core Program "International Network of Planetary Sciences." D.D. was supported by the NASA Hayabusa2 Participating Scientist Program (NNX16AL34G), and the Solar System

Exploration Research Virtual Institute 2016 (SSERVI16) Cooperative Agreement (NNH16ZDA001N) SSERVI-TREX. P.M. acknowledges funding from the French space agency CNES, from Academies of Excellence: Complex systems and Space, environment, risk, and resilience, part of the IDEX JEDI of the Université Côte d'Azur, and from the European Union's Horizon 2020 research and innovation programme under grant agreement No 870377 (project NEO-MAPP).

# Appendix. A Images of C-series bright boulders

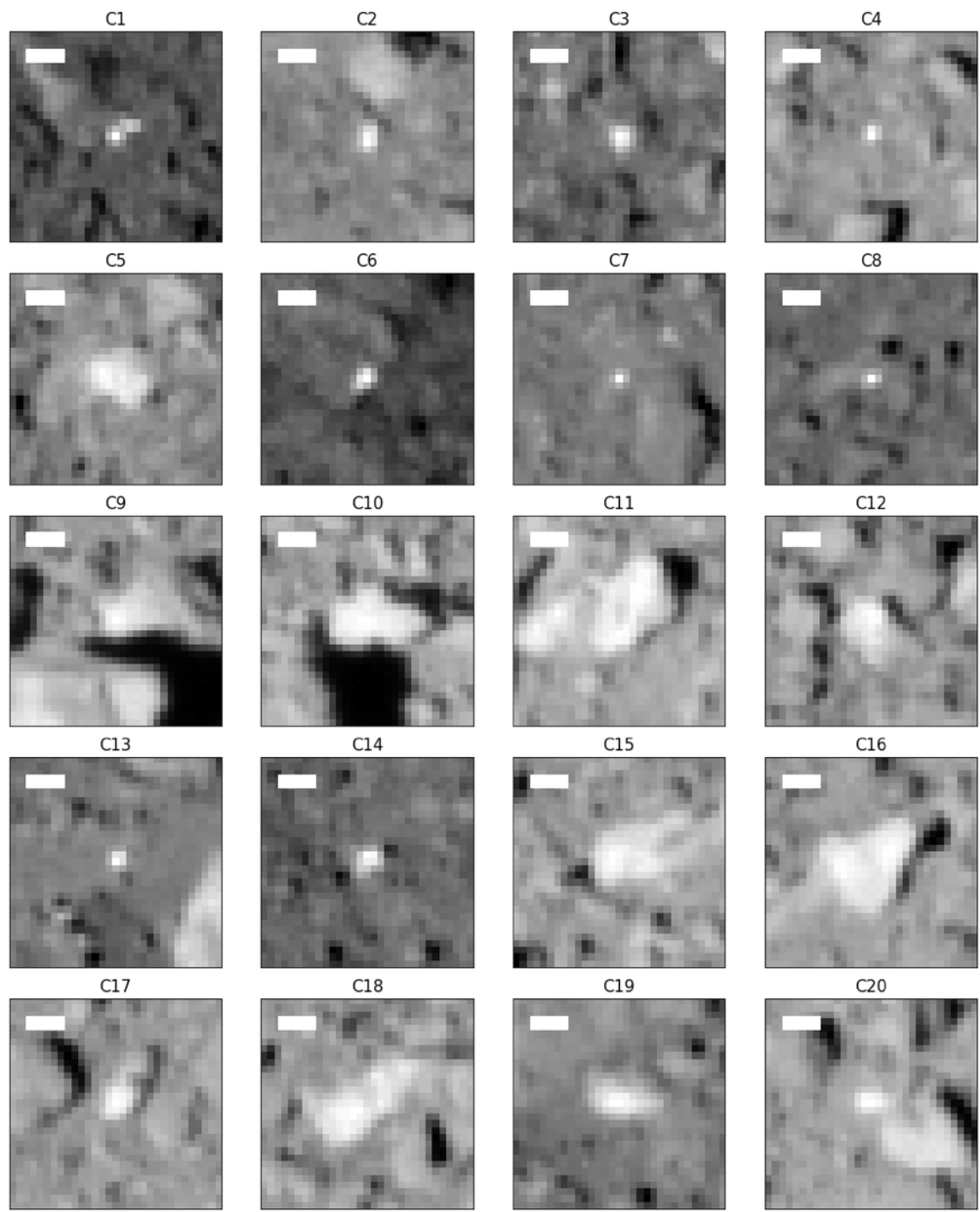

**Fig. A1.** Bright boulders observed during CRA1/CRA2 at ~1.7 km on March 21 and April 25, 2019. Size bars are 1 m. The sizes and locations of these bright boulders (C1-C79) are given in Table 1.

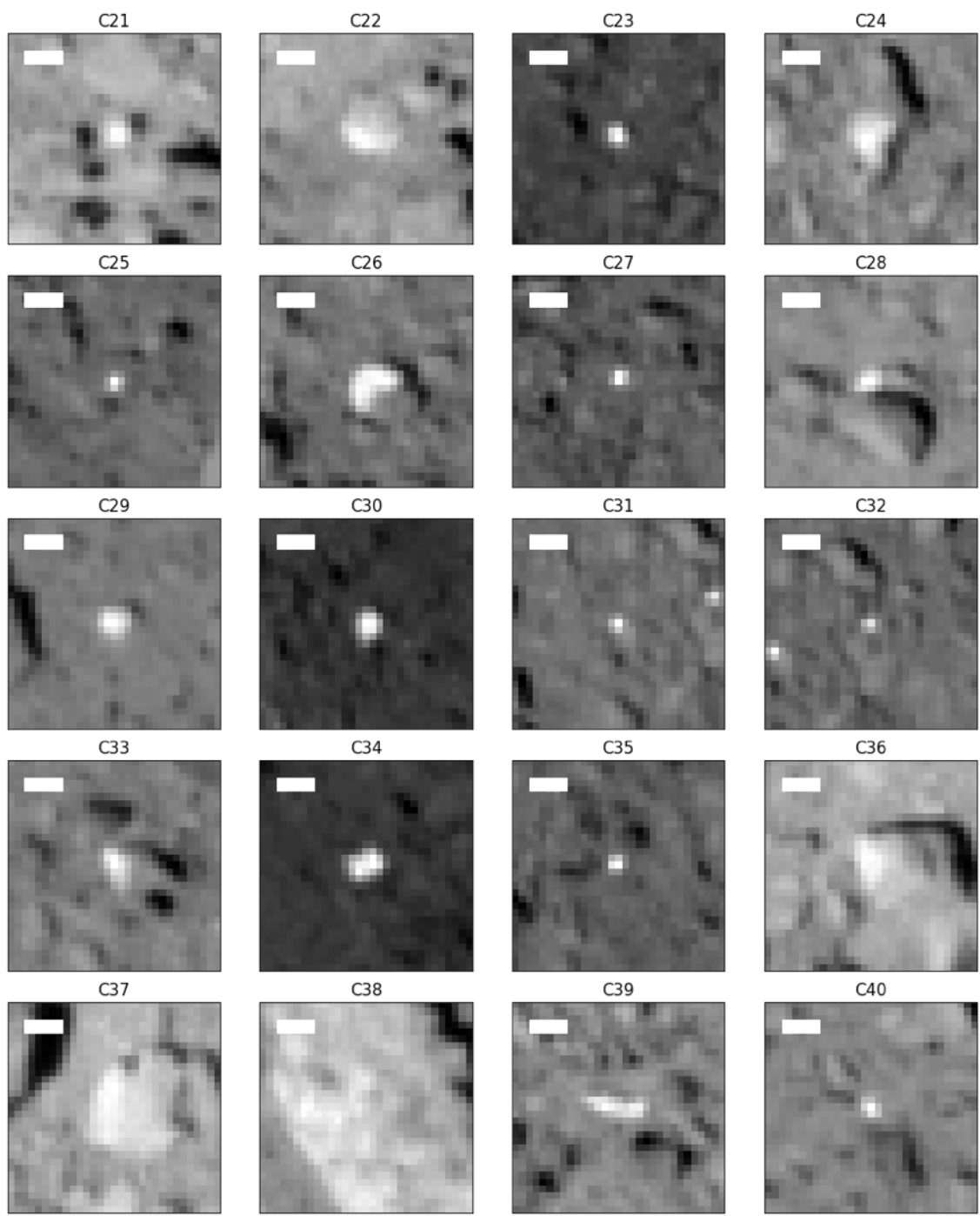

**Fig. A1.** (continued)

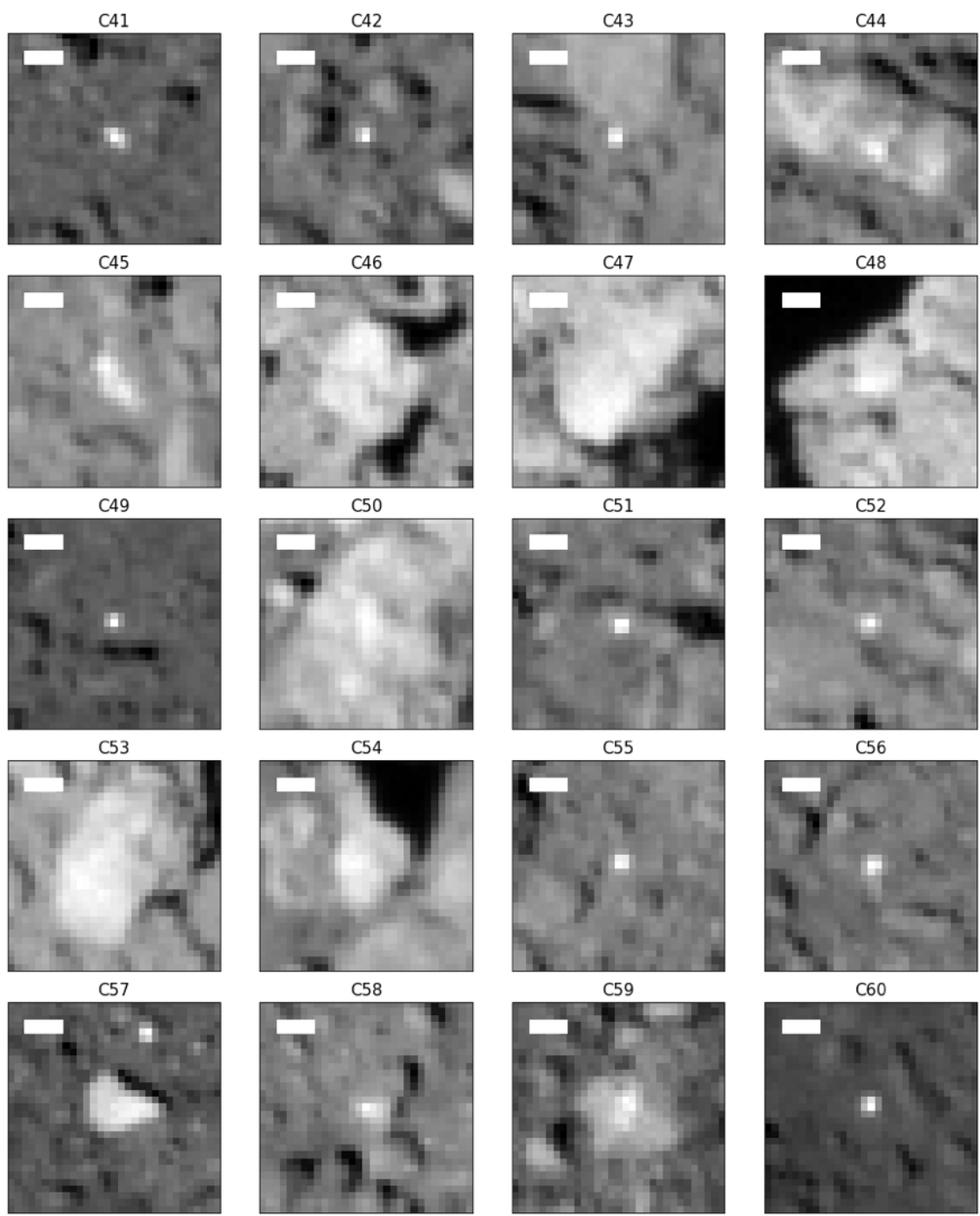

**Fig. A1.** (continued)

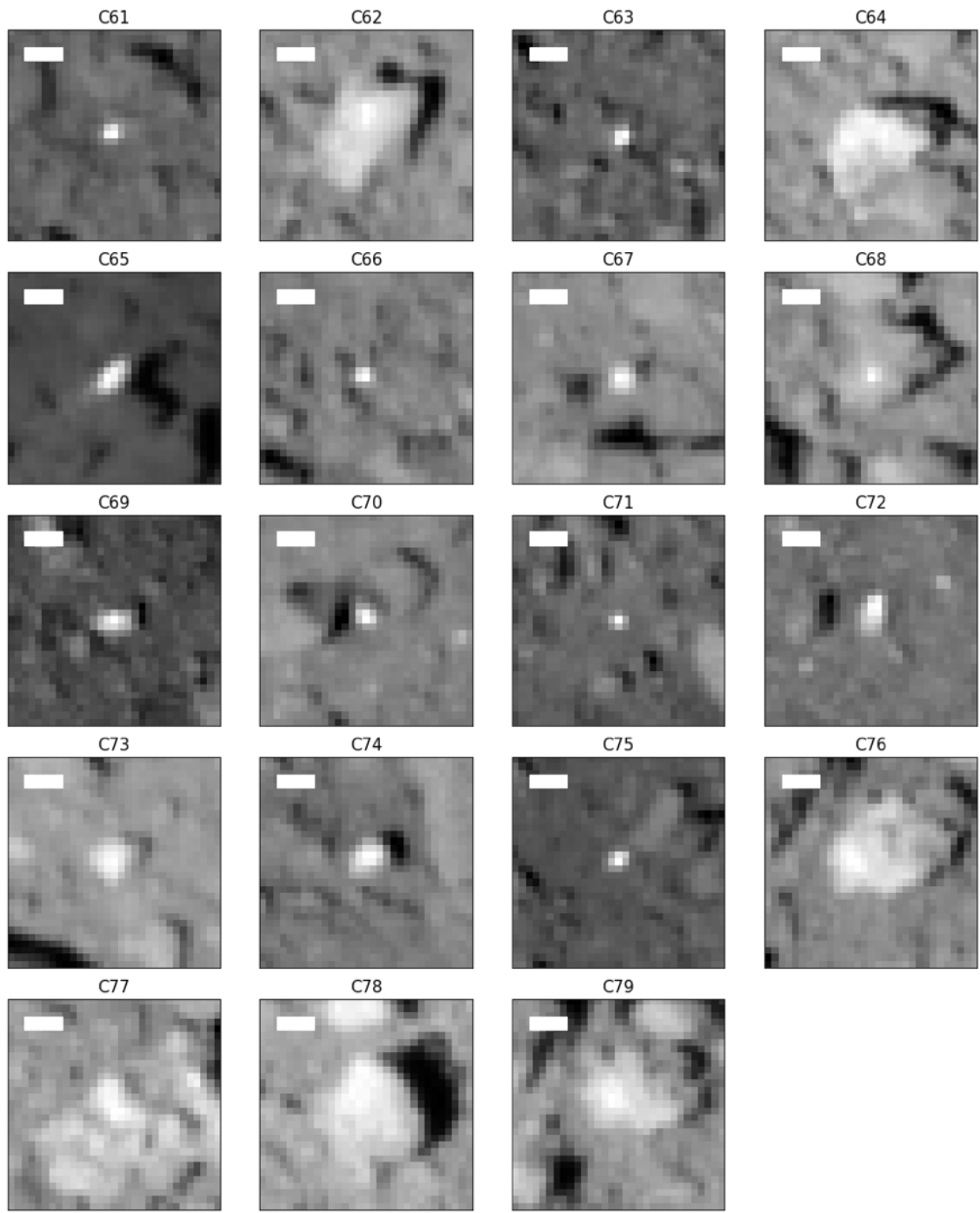

**Fig. A1.** (continued)

# Appendix. B Bright boulders detected in SFD analysis

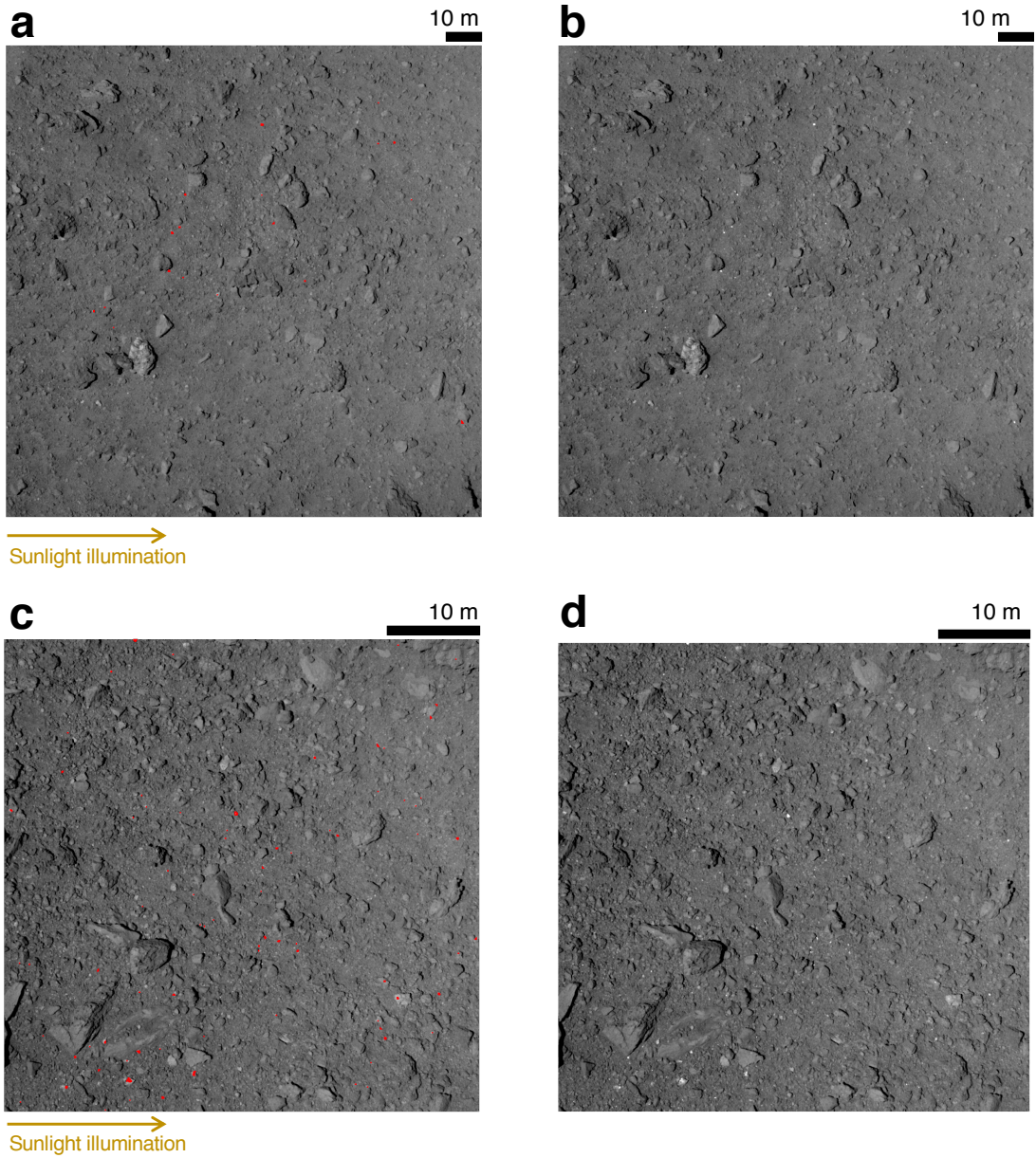

**Fig. B1.** (a, b) hyb2_onc_20181015_141813, (c,d) hyb2_onc_20181015_123017.

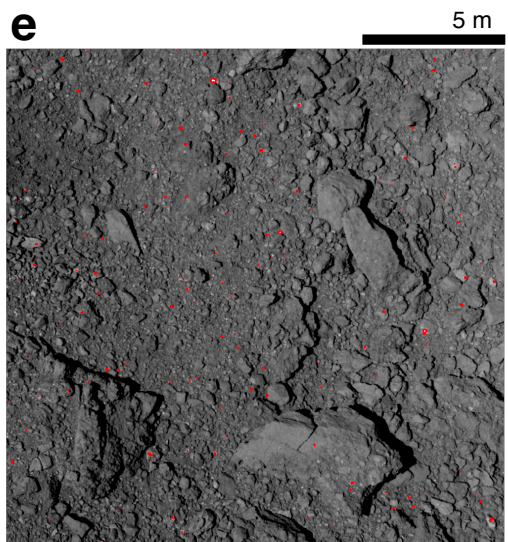 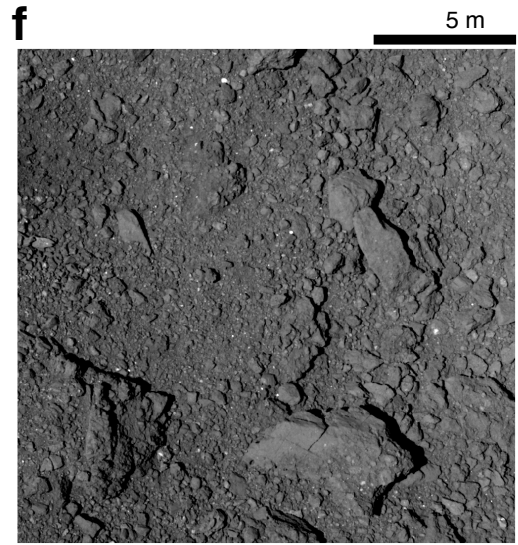
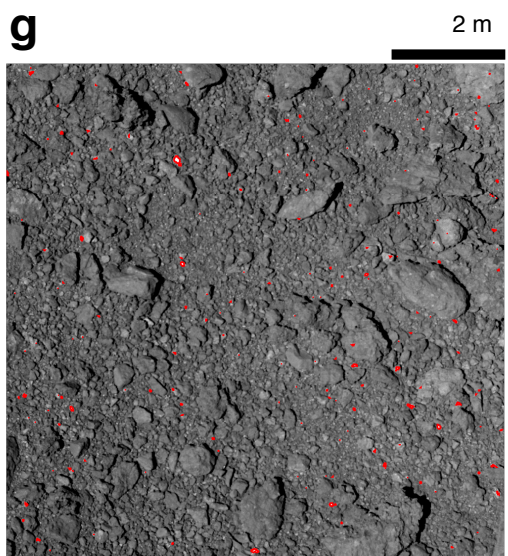 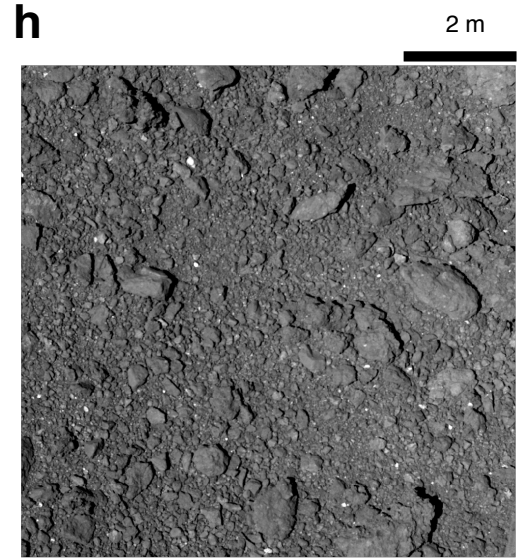

**Fig. B1. cont.** (e, f) hyb2_onc_20181015_132201, (g, h) hyb2_onc_20181015_133417.